\documentclass[12pt, hidelinks]{article}
\usepackage[small, bf]{caption}

\usepackage[utf8]{inputenc}
\usepackage{fullpage}
\usepackage{appendix}
\usepackage{graphicx}
\usepackage{amsmath}
\usepackage{pgfplots}
\usepackage{amssymb}
\pgfplotsset{compat=1.17}
\usepackage{amsthm, amsfonts}
\usepackage{url}
\usepackage{mathtools}
\usepackage{chngcntr}
\usepackage{nicefrac}
\usepackage{hyperref}
\usepackage{subdepth}
\usepackage{forest}
\usepackage{tikzducks}
\usepackage{hhline}
\usepackage{subcaption}
\usetikzlibrary{shapes.geometric,positioning}
\usetikzlibrary{decorations.pathmorphing,shapes}
\tikzset{elli/.style={ellipse,draw}}
\newsavebox\Car
\newsavebox\Tree

\usepackage{comment}
\usepackage{bbm}

\usepackage[style=alphabetic,backend=bibtex]{biblatex}
\bibliography{bibliography}

\theoremstyle{definition}

\counterwithin*{lemma}{subsection}

\usepackage{tikz}
\usepackage{subcaption}
\usetikzlibrary{arrows,automata,calc}

\tikzset{
  treenode/.style = {align=center, inner sep=0pt, text centered,
    font=\sffamily},
  arn_n/.style = {treenode, circle, white, font=\sffamily\bfseries, draw=black,
    fill=black, text width=1.5em},% arbre rouge noir, noeud noir
  arn_r/.style = {treenode, circle, red, draw=red, 
    text width=1.5em, very thick},% arbre rouge noir, noeud rouge
  arn_x/.style = {treenode, rectangle, draw=black,
    minimum width=0.5em, minimum height=0.5em}% arbre rouge noir, nil
}

\newcommand{\reals}{{\mbox{\bf R}}}

\newcommand{\eg}{{\it e.g.}}
\newcommand{\ie}{{\it i.e.}}

\newcommand{\BEAS}{\begin{eqnarray*}}
\newcommand{\EEAS}{\end{eqnarray*}}
\newcommand{\BEA}{\begin{eqnarray}}
\newcommand{\EEA}{\end{eqnarray}}
\newcommand{\BEQ}{\begin{equation}}
\newcommand{\EEQ}{\end{equation}}
\newcommand{\BIT}{\begin{itemize}}
\newcommand{\EIT}{\end{itemize}}

\title{Optimal Routing in the Presence of Hooks: \\ Three Case Studies}

\author{Tarun Chitra\footnote{The authors are listed in alphabetical order.}\\ Gauntlet \\\texttt{\small tarun@gauntlet.xyz}
    \and Kshitij Kulkarni\\ UC Berkeley\\\texttt{\small ksk@eecs.berkeley.edu} \and Karthik Srinivasan \\ Sorella Labs\\\texttt{\small karthik@sorellalabs.xyz}}

\begin{document}
\maketitle 
\begin{abstract}
We consider the problem of optimally executing a user trade over networks of constant function market makers (CFMMs) in the presence of \emph{hooks}.
Hooks, introduced in an upcoming version of Uniswap, are auxiliary smart contracts that allow for extra information to be added to liquidity pools.
This allows liquidity providers to enable constraints on trades, allowing CFMMs to read external data, such as volatility information, and implement additional features, such as onchain limit orders.
We consider three important case studies for how to optimally route trades in the presence of hooks: 1) routing through limit orders, 2) optimal liquidations and time-weighted average market makers (TWAMMs), and 3) noncomposable hooks, which provide additional output in exchange for fill risk.
Leveraging tools from convex optimization and dynamic programming, we propose simple methods for formulating and solving these problems that can be useful for practitioners.  
\end{abstract}

\section{Introduction}
Constant Function Market Makers (CFMMs) are some of the most used onchain application for trading on blockchains.
These markets, pioneered by Uniswap, allow for permissionless deployment of liquidity pools that can be used for swapping multiple cryptocurrency assets.
Uniswap has facilitated over \$2 trillion of trading volume across a number of Ethereum-compatible blockchains such as Ethereum, Arbitrum, and Optimism~\cite{uni-volume}.
The popularity of these markets has persisted throughout multiple market cycles as CFMM designs (such as Uniswap V3 \cite{adams2021uniswap}) have continued to improve.

CFMMs coordinate three principal actors: liquidity providers (LPs), arbitrageurs, and users.
Liquidity providers have passive capital and aim to earn yield on it by charging fees to users who trade against their passive capital.
For instance, if a liquidity provider owns ETH and USDC, they can facilitate trades from users between these two assets (and charge a fee in the tendered asset).
Arbitrageurs synchronize prices between centralized and decentralized venues and are crucial for ensuring CFMMs quote accurate prices~\cite{angeris2021analysis, angeris2022optimal}.
Finally, users aim to trade one asset for another for a fee.

The price charged by a CFMM to users who swap assets against the pool is controlled by an invariant (such as the constant product function used in Uniswap), which allows for computationally efficient trading.
Invariants are computed as a function of liquidity provider positions, which represent the assets locked into a smart contract by liquidity providers that are used to facilitate trades and earn fees.
Newer protocols such as Uniswap V3~\cite{adams2021uniswap} allow for users to specify prices at which their liquidity can be used in a mechanism known as concentrated liquidity, giving liquidity providers more granular control over how they earn fees.

\begin{figure}
    \begin{centering}
    \includegraphics[scale=0.3]{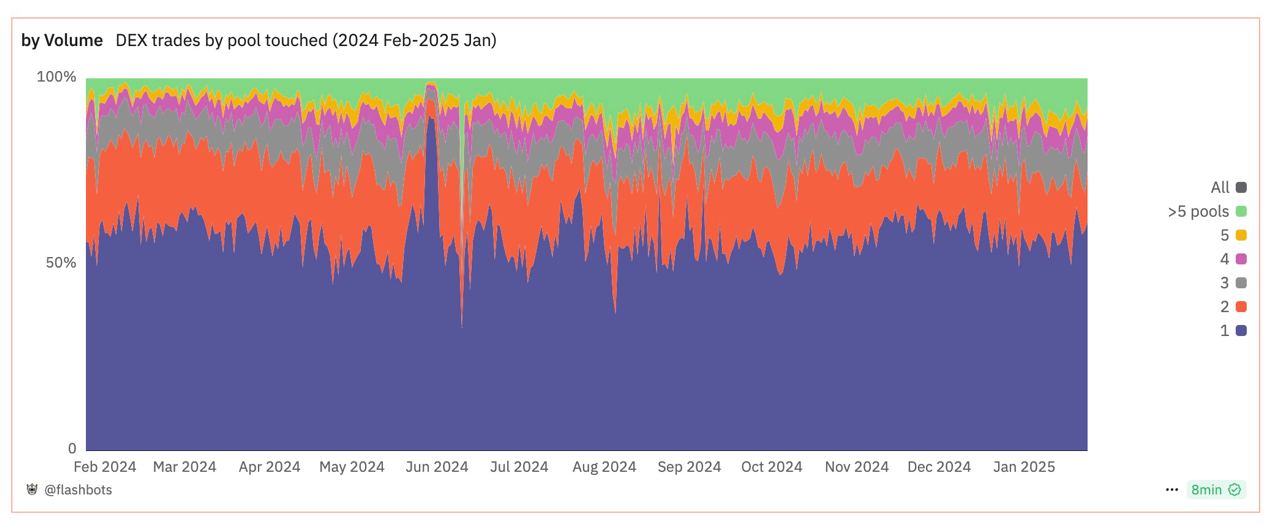}
    \caption{Decentralized exchange trades by number of pools touched~\cite{danning}.}
    \label{pools-touched}
    \end{centering}
\end{figure}

The computational efficiency of using invariants (as opposed to order books~\eg~\cite{milionis2023complexity, chitra2021liveness}) allows for easy routing across multiple pools, where a user who aims to trade asset A for asset C can efficiently route their orders via trades from A to B to C as easily as a trade directly from A to C.
Routing problems for CFMMs were introduced in~\cite{angeris2022optimal}, and were shown to admit a convex optimization formulation.
An efficient algorithm for solving the routing problem was provided in \cite{diamandis2023efficient}, based on a decomposition method.
Together, these results demonstrate that the \emph{optimal} routing of user trades through networks of CFMMs is \emph{efficient}.
Nearly 50\% of all Uniswap trades are routed over multiple pools via services like Uniswap Router, Matcha, 1inch, and Metamask (see Figure~\ref{pools-touched}), demonstrating the importance of routing to the user experience and growth of CFMMs.

\paragraph{Hooks.}
Hooks, introduced in Uniswap v4, allow anyone to deploy concentrated liquidity pools with custom functionality \cite{uniswapv4}.
A goal of hooks is to allow application developers to add custom rules that modify the output quantity of assets tendered to a swapper as a function of other blockchain state.
This modification can be viewed as a change to the output of a CFMM that is dependent on additional data beyond the trade size and reserves.
Hooks operate via smart contracts that interface with a pool at various stages in its lifetime.
For example, they may be used to set dynamic swap fees relative to the volatility of an asset, implement a time-weighted average market maker for large trades, or allow for onchain limit orders \cite{hookswebsite}.
These use-cases aim to provide utility to both swappers and LPs that is distinct from the one specified by the invariant. 

Mathematically, one can view hooks as allowing liquidity providers to realize different payoffs than those replicable by CFMMs (\eg~replicating market makers~\cite{angeris2023replicating}) by adding additional data to trades. This also means that routers, who compute optimal paths through these pools, are able to respond to these changing payoffs. The presence of hooks therefore necessitates a reexamination of the kinds of methods that routers typically use to provide swappers with optimal routing. 

Routing in the presence of hooks is nontrivial and opens up a new design space. Consider the example of onchain limit orders. Initial versions of the Uniswap protocol only allowed for the use of the reserves deposited by the liquidity providers. This meant that routers were constrained to using this deposited liquidity across all pools. If onchain limit orders can be implemented via hooks, routers have access to new `standing liquidity' that can be used to improve swapper outcomes. Further, hooks can use additional external information, such as information about the volatility of an asset, which can inform the strategies that routers use to route trades over pools and over time. This can be valuable for improving swapper outcomes by strategically trading in times that are favorable to the user.

\paragraph{This paper.}
Motivated by this new design space, we initiate a study of optimal routing in the presence of hooks and provide three case studies for practical hooks that affect routers: \textbf{1)} onchain limit orders, \textbf{2)} optimal liquidations that use volatility information to improve user outcomes, and \textbf{3)} noncomposable hooks, which are pools that provide additional output due to external market makers in exchange for fill risk. In each of these cases, we formulate optimization problems that can be solved using convex optimization or dynamic programming methods, and provide insight into the solutions of these problems.

First, we provide an introductory background on CFMMs and optimal routing in Section \ref{sec:background}.
Subsequently, in Section \ref{sec:limit-orders}, we study optimal routing through limit orders. We introduce the notion of the \emph{trading set} of a limit order, which is the set of all trades a router can make that are feasible for the limit order. Further, we show that limit orders act as concentrated liquidity positions at infinitesimally small price ticks. These definitions allow us to modify the optimal routing problem in \cite{angeris2022optimal} to account for standing liquidity provided by arbitrarily many limit orders. The resulting optimization problem remains convex, and we demonstrate numerical results for a simple two-asset network and for a general network of many markets and assets. 

In Section \ref{sec:temporal-hook}, we consider hooks that route trades over time, and specifically look at the example of a user attempting to liquidate a large amount of an asset into another asset on a CFMM. We consider two strategies. First, we consider the time-weighted automated market maker (TWAMM), which allows the user to split the trade uniformly over a time horizon in a gas-efficient manner.
We compare this type of market maker to an optimal liquidation problem when the user has access to a mispricing signal relative to an external market, and can trade when the mispricing is favorable, in exchange for paying gas every trade.
This problem is formulated as a Markov decision process (MDP), which we solve numerically using dynamic programming methods.
We demonstrate that the optimal liquidation strategy can outperform the TWAMM strategy of uniformly splitting the trade when the asset is sufficiently volatile. 

Finally, in Section \ref{sec:non-composable-hooks}, we look at noncomposable hooks, which can implement sovereign liquidity pools that have control over their state. These pools cannot be routed through in a manner that is composable with conventional CFMMs, but provide an opportunity for external market makers to provide additional liquidity at preferential prices. We model
and quantify the tradeoff between this preferential execution and the nondeterministic execution risk that users face on these hooks. We cast noncomposable hook routing as a mean-variance optimization problem, and show the efficient frontier for a user's trade. 

Our goal in this paper is to demonstrate that simple methods can allow hooks to be easily incorporated in existing routing schemes when the corresponding routing problem is well-structured. Code for all numerical experiments in this paper can be found at this link: 
\begin{center}
    \url{https://github.com/kkulk/hooks}.
\end{center}

\section{Background}\label{sec:background}
In this section, we provide a brief background on CFMMs and the optimal routing problem for completeness.
We refer the reader to the papers~\cite{angeris2022constant, angeris2020improved, angeris2022optimal, angeris2022does, angeris2023geometry} for a more thorough coverage of the preliminaries.

\paragraph{Constant function market makers.}
A \emph{constant function market maker} (CFMM) is a contract that holds some amount of \emph{reserves} $R, R' \ge 0$ of two assets, $A$ and $B$,
and has a \emph{trading function} $\varphi: \reals^2\times\reals^2 \to \reals$.
Users can submit a \emph{trade} $(\Delta, \Delta')$ denoting the amount of asset $A$ they would like to tender (if $\Delta$ is positive) to receive $\Delta'$ units of asset $B$.
The contract accepts the trade only if $\varphi(R, R', \Delta, \Delta') = \varphi(R, R', 0, 0)$,
and pays out $\Delta'$ to the user, which is implicitly defined by the trading function.

\paragraph{Forward exchange function.} 
We briefly summarize the main definitions and results of~\cite{angeris2022does} here, which provide a characterization of the output the user receives from a CFMM as a function of the trading function.
Suppose that the trading function $\varphi$ is differentiable (as most trading functions in practice are). Then
the \emph{forward exchange rate} for a trade of size $\Delta$ is
$
g(\Delta) = \frac{\partial_3 \psi(R, R', \Delta, \Delta')}{\partial_4 \psi(R, R', \Delta, \Delta')}.
$
Here $\partial_i$ denotes the partial derivative with respect to the $i$th argument,
and $\Delta'$ is specified by the implicit condition $\psi(R, R', \Delta, \Delta') = \psi(R, R', 0, 0)$; \ie, the trade
$(\Delta, \Delta')$ is assumed to be valid. Additionally, the reserves $R, R'$ are assumed to be fixed.
The function $g$ represents the marginal forward exchange rate of a positive-sized trade with the CFMM. One important property of $g$ is that it can be used to compute $\Delta'$~\cite[Section 2.1]{angeris2022does} by
\begin{align*}
    \Delta' = \int_{0}^{-\Delta} g(t) dt.
\end{align*}
We define $G(\Delta) = \Delta'$ to be the \emph{forward exchange function}, which is the amount of output asset received for an input of size $\Delta$. Whenever we reference the function $G(\Delta)$ for a given CFMM, we always make clear the reserves associated with that CFMM, or the instantaneous price it quotes along with its liquidity parameter, as defined in \cite{adams2021uniswap}. We note that $G(\Delta)$ was shown to be concave and increasing in \cite{angeris2022constant}, with $G(0) = 0$. This function also takes into account the fee charged by the CFMM. 

\paragraph{Routing.} The problem of routing trades through a network of CFMMs was studied in \cite{angeris2022optimal}, in which the optimal routing problem was shown to be convex.
Given a network of $m$ CFMMs trading $n$ assets, the optimal routing problem can be formulated as
\begin{align*}
    \text{maximize} \quad & U(\Psi) \\
    \text{subject to} \quad & \Psi = \sum_{i=1}^{m} A_i \Delta_i, \\
    \quad & \Delta_i \in T_i,
\end{align*}
with variables $\Psi \in \mathbf{R}^n$ and $\Delta_i \in \mathbf{R}^{n_i}$ for $i = 1, \dots, m$ and a concave utility function $U : \reals^n \rightarrow \reals$.
The matrices $A_i \in \mathbf{R}^{n \times n_i}$ are so-called local-global matrices that convert the indices of assets in each CFMM to a global index.
The utility function $U$ encodes the user's satisfaction with a net trade $\Psi$ with the network.
The sets $T_i$ are the \emph{trading sets} of each CFMM, and encode the feasible trades that each market can tender to the user.
It was shown in \cite{angeris2020improved} that the trading sets can be written using the trading function of the corresponding CFMM by
\begin{align*}
    T_i = \{(z^1, z^2) \in \mathbf{R}^{n_i} \times \mathbf{R}^{n_i} | \varphi_i(R_i + \gamma_i z^1 - z^2) \geq \varphi_i(R_i) \},
\end{align*}
where $\gamma_i \in [0,1]$ is the fee parameter of the CFMM. Because $\varphi_i$ are concave functions, the sets $T_i$, which are superlevel sets of concave functions, are convex. An efficient algorithm for solving this problem based on a decomposition method was given in \cite{diamandis2023efficient}. Several other examples of utility functions involving arbitrage and liquidation of a basket of assets were provided in \cite{angeris2022optimal}.

\section{Routing Through Limit Orders}\label{sec:limit-orders}
In this section, we describe how the original CFMM routing problem can be modified to route through limit orders, which are orders that give users guarantees on the maximum (minimum) prices the user receives on buy (sell) orders.
Early versions of Uniswap and other CFMMs only allowed routers to make use of the liquidity provided in the core protocol; hooks enable the implementation of onchain limit orders.
Consequently, our formulation allows the routing problem to use limit orders such as `standing liquidity' that can be incorporated with traditional sources of liquidity provided by CFMMs.
Limit orders can be used by routers in a capital-efficient manner to improve the end user's experience, and are examples of hooks that modify the user's output with auxiliary data that can be attached to the routing problem.

We first show how limit orders can be viewed through the lens of \emph{trading sets}, and how they modify the forward exchange function when composed with a CFMM.
Further, limit orders can be viewed as `concentrated liquidity' positions on a price tick of infinitesimally small width. With this background, we show how the optimal routing problem can be modified to use liquidity provided by limit orders in conjunction with CFMMs to improve the user's output; the problem remains a convex optimization problem and we present efficient numerical methods that can be used to solve it. 

\subsection{Limit Orders}
\begin{figure}[t]
    \centering
    \begin{tikzpicture}
        \begin{axis}[
            width=10cm, height=8cm,
            xlabel={Input, $z_1$},
            ylabel={Output, $z_2$},
            xmin=0, xmax=10,
            ymin=0, ymax=10,
            domain=0:10,
            samples=100,
            legend pos=north east,
            axis x line=bottom,
            axis y line=left,
            enlargelimits=false,
            grid=major
        ]
            % Feasible region shading
            \addplot [
                fill=blue!30, draw=none, opacity=0.7
            ] {min(2, 0.5*x)} \closedcycle;

            % Price constraint line: z2 = p0 * z1 (p0 = 0.5) in green dashed
            \addplot [
                domain=0:10,
                thick, green, dashed
            ] {0.5*x};
            \addlegendentry{$z^2 = p_0 z^1$}

            % Value constraint line: z2 = V0 (V0 = 2) in red dashed
            \addplot [
                domain=0:10,
                thick, red, dashed
            ] {2};
            \addlegendentry{$z^2 = V_0$}
        \end{axis}
    \end{tikzpicture}
    \caption{Trading set, $\tilde{T} = \{z \in \mathbf{R}^2 \mid z^2 \leq p_0 z^1, z^2 \leq V_0, z^1, z^2 \geq 0\}$, for $p_0= 0.5$, $V_0 = 2$.}
    \label{fig:feasible-set-limit}
\end{figure}
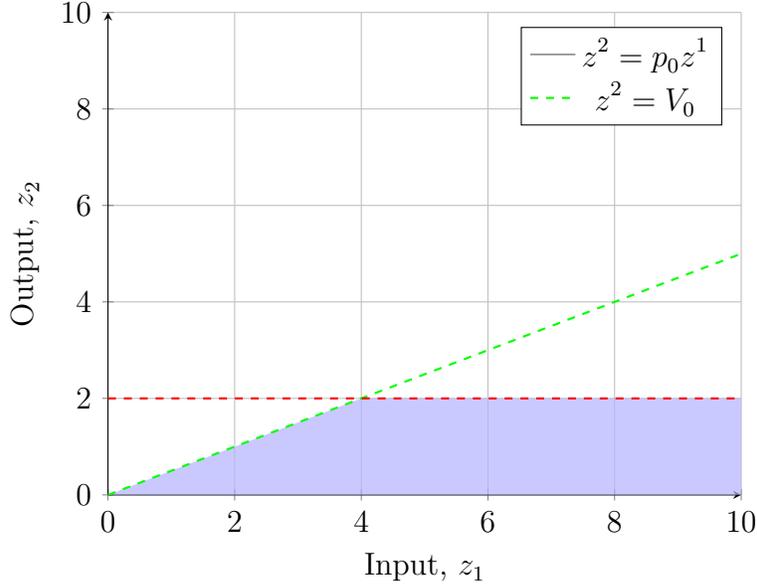

\paragraph{The trading set of a limit order.}
A limit buy (sell) order is an agreement to exchange a volume $V_0$ of an asset  $B$ for an asset $A$ for no more (less) than a price $p_0$. Trading with a limit order as a counterparty means tendering an amount of an asset to the user that satisfies the price and volume constraints. Formally, while trading against a limit buy order, we can define a variable $z \in \mathbf{R}^2$, where $z_1$ is the amount of asset $A$ tendered and $z_2$ is the amount of asset $B$ received. In this spirit, we define the \emph{trading set of a limit (buy) order} as 
\begin{align*}
    \tilde{T} = \{z = (z^1, z^2) \in \mathbf{R}^2 \ \rvert \ p_0 z^1 - z^2 \geq 0, z^2 \leq V_0, z^1, z^2 \geq 0\}.
\end{align*}
In words, this defines all the trades that satisfy the volume and price constraints of the limit buy order. A corresponding set can be defined for sell orders. In general this is a trapezoidal set, and is a convex set because it is the intersection of half-planes. We depict an example in Figure \ref{fig:feasible-set-limit}. We will shortly that the trading set of a limit order is a natural object to consider from the perspective of optimal routing.

\paragraph{Composition rule.} There is a simple composition rule for multiple limit orders that trade between the same assets. Suppose there are two limit buy orders that trade between assets $B$ and $A$, with respective trading sets $\tilde{T}_1$ and $\tilde{T}_2$. The joint trading set of the two orders is the Minkowski sum of the two trading sets, i.e.
\begin{align*}
    \tilde{T} = \{ z = z_1 + z_2 | z_1 \in \tilde{T}_1 , z_2 \in \tilde{T}_2\}.
\end{align*}
Because the Minkowski sum of convex sets is also convex, $\tilde{T}$ remains a convex set. This composition rule can be naturally extended to arbitrarily many limit orders. Therefore, without loss of generality, at any time, we can associate to a pair of assets a single convex set that encodes the constraints of all the limit orders in that pair. 
% \begin{figure}[t]
%     \centering
% \includegraphics[width=0.65\linewidth]{figures/limit-order/feasible-set-limit.png}
%     \caption{The feasible set of a limit order; $V_0 = 2$, $p_0 = 0.5$.}
%     \label{fig:feasible-set-limit}
% \end{figure}

\begin{figure}[t]
\centering
\begin{tikzpicture}
    % Set up the axes
    \draw[->] (-0.5,0) -- (7,0) node[right] {Input};
    \draw[->] (0,-0.5) -- (0,5) node[above] {Output};
    
    % Plot the original constant product curve
    \draw[blue!30, thick] plot [domain=0:6.5] (\x,{4 - 4/(1 + \x)});
    
    % Plot the modified curve with limit order
    \draw[blue, thick] plot [domain=0:1] (\x,{4 - 4/(1 + \x)});  % First segment
    \draw[blue!20, thick] plot [domain=1:2] (\x,{\x+1});  % Linear segment (limit order)
    \draw[blue, thick] plot [domain=2:6.5] (\x,{4 - 4/(1 + (\x-1))+1});  % Final segment
    
    % Add labels and annotations
    \node[below] at (1,-0.2) {$\Delta_1$};
    \node[below] at (2,-0.2) {$\Delta_2$};
    
    % Add legend
    \node[blue!30, right] at (5,3) {without limit order};
    \node[blue, right] at (4,4.7) {with limit order};
    
    % Add price slope indicator
    \draw[red, dashed] (1,2) -- (2,3) node[above left] {slope=$p_0$};

        % Add vertical dashed lines at transition points
    \draw[gray!35, dashed] (1,0) -- (1,5);
    \draw[gray!35, dashed] (2,0) -- (2,5);

    % \draw[decorate,decoration={brace,amplitude=6pt,mirror}] (1,-0.8) -- (2,-0.8);
    % \node[below] at (1.5,-1) {$V_0$};
\end{tikzpicture}
\caption{Comparison of the forward exchange function of a CFMM with and without a limit order. The output with the limit order creates a linear segment at the limit price $p_0$, followed by a continuation of the CFMM's forward exchange function.}
\label{fig:modified-forward-exchange}
\end{figure}
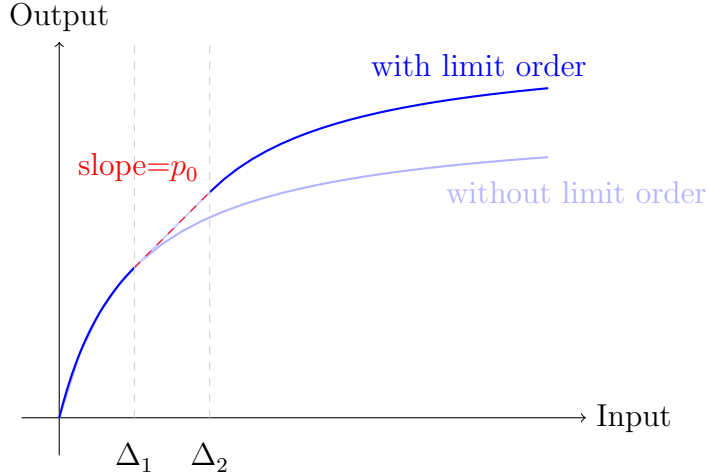

\paragraph{Modified forward exchange function.} A limit order can be combined with a standard CFMM, with a forward exchange function $G(\Delta)$, to construct a modified forward exchange function $\tilde{G}(\Delta)$ which is a function of the auxiliary data $(p_0, V_0)$.
The modified function can be defined piecewise, with three regions dictated by the price and volume the limit order. Define $\Delta_1$ to be the input amount where the forward exchange function of the CFMM first reaches the limit price $p_0$, such that $\lim_{\Delta \rightarrow \Delta_1^-} G'(\Delta) = p_0$, and $\Delta_2$ to be the point where the limit order is exhausted. Then, the modified function (depicted in Figure \ref{fig:modified-forward-exchange}) can be defined as
\begin{align*}
    \tilde{G}(\Delta) = \begin{cases}
        G(\Delta) & \text{if } \Delta \leq \Delta_1 \\
        G(\Delta_1) + p_0(\Delta - \Delta_1) &\text{if } \Delta_1 < \Delta < \Delta_2 \\
        G(\Delta- (\Delta_2-\Delta_1)) + p_0(\Delta_2-\Delta_1) &\text{if } \Delta \geq \Delta_2
    \end{cases}
\end{align*}

Despite being constructed piecewise, the modified forward exchange function is differentiable throughout its domain. Differentiability follows from two key properties. First, the function is continuous across its entire domain. Second, while the function consists of three distinct components, each component is differentiable on its respective subdomain, and at the transition points $\Delta_1$ and $\Delta_2$, the left and right derivatives coincide, both equaling the limit price $p_0$. The equality of directional derivatives ensures that the derivative is well-defined at these boundary points. 

The modified forward exchange function is also concave over its domain. On the intervals $[0,\Delta_1]$ and $[\Delta_2,\infty)$, the function inherits the concavity of the forward exchange function of the CFMM. On the interval $[\Delta_1,\Delta_2]$, the function is linear with slope $p_0$, and is thus weakly concave. Finally, at the transition points $\Delta_1$ and $\Delta_2$, the first derivative (which equals $p_0$ at these points) is non-increasing when passing from one region to the next. Together, these properties show the concavity of $\tilde{G}(\Delta)$, preserving the economic property that forward exchange rates are non-increasing in trade size even in the presence of limit orders.

\paragraph{Limit orders and concentrated liquidity.} Limit orders have an interpretation from the perspective of concentrated liquidity positions. The liquidity of a CFMM represents the degree to which the market is resistant to a change in price at a particular price level. When trading against a limit order, the market remains infinitely resistant to a change in price until the order's volume is cleared, which gives us our interpretation of the limit order as a concentrated liquidity position at an infinitesimally small price tick. 

Following Uniswap V3's \cite{adams2021uniswap} notation, consider liquidity concentration over a price tick $[p_a, p_b]$. The liquidity distribution in Uniswap V3 can be represented as a step function,
\begin{align*}
L(p) = \begin{cases}
L_{(a,b)} & \text{if } p \in [p_a, p_b] \\
0 & \text{otherwise}
\end{cases}
\end{align*}
where $L_{(a,b)}$ is the liquidity provided in the tick, derived from the tick's reserves and trading function. A limit order can be viewed as the extreme case of concentrated liquidity where all liquidity is concentrated at a single price point $p_0$. Mathematically, we can represent this as the limit of a sequence of increasingly concentrated liquidity positions, converging to a scaled Dirac delta function at the limit price. 

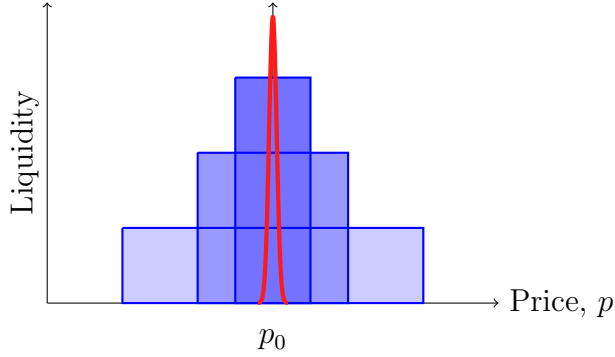
\begin{figure}[t]
\centering
\begin{tikzpicture}
% Define common styles for different opacities of blue
\tikzstyle{plot1}=[blue!20, thick, fill=blue!20]
\tikzstyle{plot2}=[blue!40, thick, fill=blue!40]
\tikzstyle{plot3}=[blue!55, thick, fill=blue!55]
\tikzstyle{delta}=[red!90, ultra thick]

% Draw axes
\draw[->] (-3,0) -- (3,0) node[right] {Price, $p$};
\draw[->] (0,0) -- (0,4);
\node[below] at (0,-0.2) {$p_0$};
\draw[->] (-3,0) -- (-3,4); 
\node[rotate=90] at (-3.3,2) {\text{Liquidity}};

% Draw reference point

% Draw increasingly concentrated liquidity positions as filled rectangles
\fill[plot1] (-2,0) rectangle (2,1);
\fill[plot2] (-1,0) rectangle (1,2);
\fill[plot3] (-0.5,0) rectangle (0.5,3);

% Draw the outlines
\draw[blue, thick] (-2,1) -- (-2,0) -- (2,0) -- (2,1) -- (-2,1);
\draw[blue, thick] (-1,2) -- (-1,0) -- (1,0) -- (1,2) -- (-1,2);
\draw[blue, thick] (-0.5,3) -- (-0.5,0) -- (0.5,0) -- (0.5,3) -- (-0.5,3);

% Draw approximation of delta function
\draw[delta] plot [domain=-0.2:0.2, samples=100] 
  (\x,{3.8*exp(-(\x*\x)/0.005)});

\end{tikzpicture}
\caption{Visualization of a limit order as a concentrated liquidity position. The sequence of blue rectangular functions with increasing height and decreasing width approaches a delta function representing a limit order at price $p_0$, with volume $V_0$. Darker shades of blue indicate more concentrated liquidity positions.}
\label{fig:concentrated-liquidity}
\end{figure}

Consider a sequence of intervals $[p_0 - \epsilon_n, p_0 + \epsilon_n]$ where $\epsilon_n \to 0$ as $n \to \infty$. For each $n$, define the liquidity distribution

\begin{align*}
L_n(p) = \begin{cases}
\frac{V_0}{2\epsilon_n} & \text{if } p \in [p_0 - \epsilon_n, p_0 + \epsilon_n] \\
0 & \text{otherwise}
\end{cases}
\end{align*}
The scaling factor $\frac{1}{2\epsilon_n}$ ensures that the total liquidity remains constant, such that
\begin{align*}
\int_{-\infty}^{\infty} L_n(p) dp = V_0.
\end{align*}
As $\epsilon_n \to 0$, this sequence converges to a scaled Dirac delta function:
\begin{align*}
\lim_{n \to \infty} L_n(p) = V_0\delta(p - p_0).
\end{align*}
This limit represents a pure limit order at price $p_0$ with volume $V_0$ and is depicted in Figure \ref{fig:concentrated-liquidity}; this maps to a concentrated liquidity position that provides infinite liquidity at $p_0$ and zero liquidity everywhere else, scaled by the volume to be cleared at the instantaneous price of the limit order.

% \paragraph{Replicating generic output functions.} \KK{check with Tarun}
% Suppose that we have a CFMM invariant with a forward exchange function $G(\Delta)$.
% We can construct a generic output modifying hook via a modified forward exchange function $\tilde{G}(\Delta; x)$, where $x \in \reals^n$ is the auxiliary data used in the hook.
% For instance, for a limit order, $x$ refers to the maximum or minimum price $p_0$ that is used to construct the order.
% Generally, since $\tilde{G}$ is non-decreasing, the inverse function $F : [0,\infty) \rightarrow \reals$ defined by $F(p) = \tilde{G}^{-1}(p;x) = \min\{ \Delta : \tilde{G}(\Delta; x) \geq p\}$ exists.
% The output of this function is the trade size needed to cause the market to reach a price $p$.

% For any second-order differentiable function $f(x) \in C^2([0, \infty))$, the Carr-Madan formula~\cite{bossu2021functional, carr1998towards} expands $f$ as
% \[
% f(x) = f(x_0) + f'(x_0)(x-x_0) + \int_{0}^{x_0} f''(y) \max(y-x, 0)dy + \int_{x_0}^{\infty} f''(y) \max(x-y, 0) dy 
% \]
% The kernels $\max(x-y, 0)$ can be thought of as limit orders.
%  To see this, note that the boundary of the trading set of a limit order is a translation and reflection of the function $f(y;x) = \max(x-y, 0)$.
% This implies that any sufficiently smooth output modifying hook with inverse forward exchange function $F(p)$ can be replicated via a (potentially infinite) linear combination of limit orders.

\subsection{The Routing Problem}\label{sec:routing-limit}
The original routing problem can now be modified straightforwardly to account for standing liquidity provided via limit orders by appending additional variables and their corresponding trading sets.

\paragraph{The modified routing problem.} The modified routing problem is simple to describe. Suppose there are $k$ limit orders. 
We introduce the limit order variables $z_j \in \mathbf{R}^2$ for each order, each constrained by their trading sets $\tilde{T}_j$, and introduce additional `local-to-global' matrices $B_j$ that convert the limit orders into net trades, giving us the optimization problem 
\begin{align*}
    \text{maximize} \quad & U(\Psi) \\
    \text{subject to} \quad & \Psi = \sum_{i=1}^{m} A_i x_i + \sum_{j=1}^{k} B_j z_j,  \\ 
    & x_i \in T_i, \\ 
    & z_j \in \tilde{T}_j,
\end{align*}
where $j = 1,\dots, k$. Notably, because the trading sets of limit orders are convex, this problem remains a convex optimization problem. Interestingly, in the absence of CFMMs, and with linear utility functions, this problem would be a linear program, because the limit order constraints are linear inequalities. Combined with CFMMs, however, the problem becomes nonlinear, but can be solved using standard convex optimization methods. 

\begin{figure}[t]
\centering
\begin{minipage}[b]{0.45\textwidth}
    \centering
    \begin{tikzpicture}
    \node[draw, circle] at (-.5,0) {$A$};
    \node[draw, circle] at (5.5,0) {$B$};
    \draw[->, >=latex] (0, 0.2) to[out=30, in=150] (5,0.2);
    \node at (2.5, 1.25) {$\varphi(R,R', x, y) = (R+x)(R'-y)$};
    \draw[->, >=latex] (0, -0.2) to[out=-30, in=-150] (5,-0.2);
    \node at (2.5, -1.25) {$(p_0,V_0)$};
    \end{tikzpicture}
    \captionof{figure}{The CFMM Pigou network with limit orders.}
    \label{fig:pigou}
\end{minipage}
\hfill
%----- Second figure in minipage -----
\begin{minipage}[b]{0.47\textwidth}
    \centering
    \includegraphics[width=\textwidth]{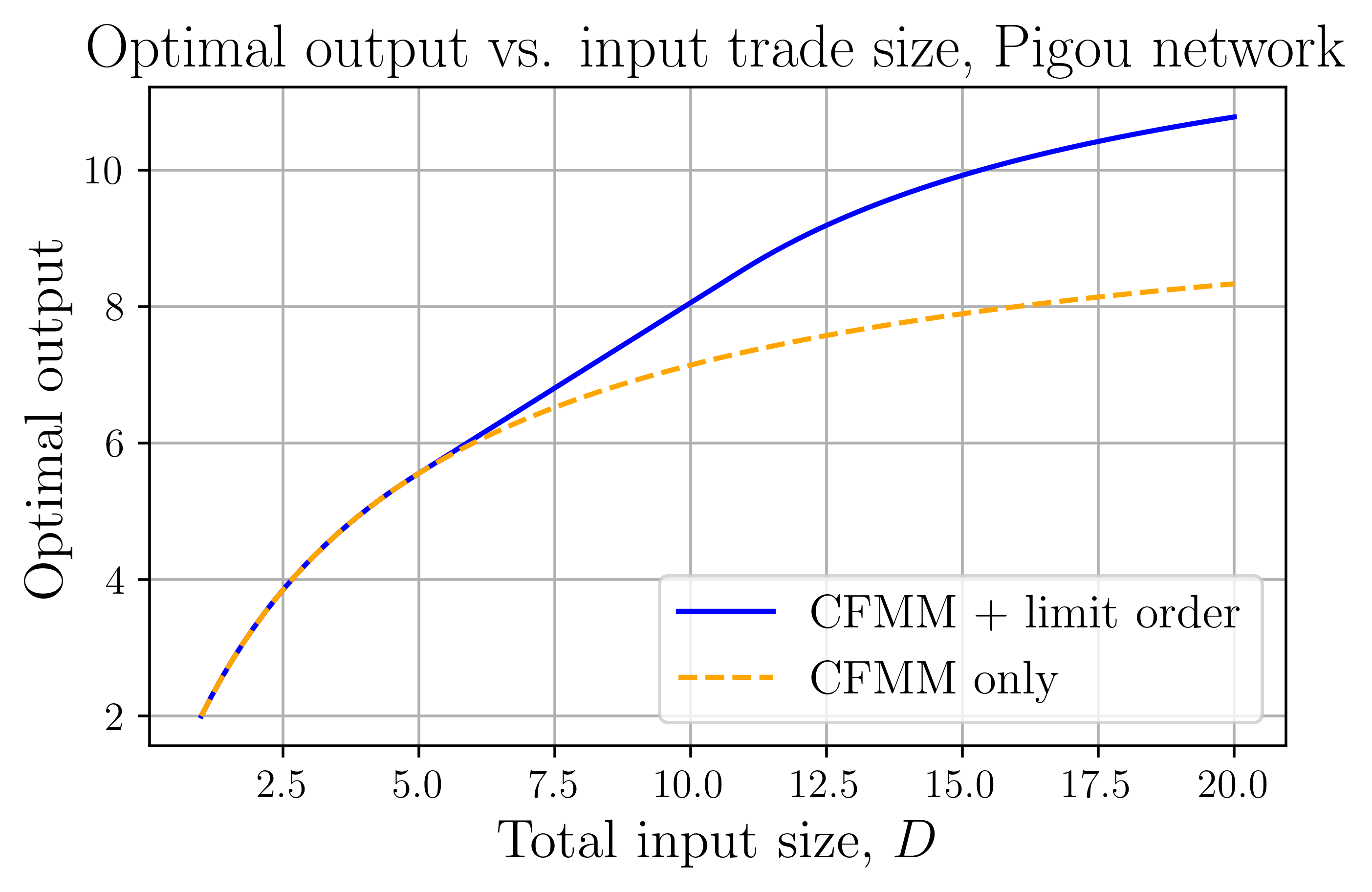}
    \captionof{figure}{Optimal output for the Pigou network with limit orders as $\Delta$ varies.}
    \label{fig:pigou-limit}
\end{minipage}
\end{figure}

\paragraph{Pigou network with limit orders.}  To see a simple example of how limit orders can be integrated into routing, we consider the Pigou network, a two-link network that was first studied in the context of CFMMs in \cite{kulkarni2023routing}. In this network, there are two assets, $A$ and $B$, and two markets trading between them, and a user wants to provide no more than $D$ units of asset $A$ to maximize the amount of asset $B$ they receive. The first market is a constant product market maker with trading function $\varphi(R, R', x,y) = (R+x)(R'-y)$, and the second market is a single limit order at price $p_0$ with volume $V_0$, as depicted in Figure \ref{fig:pigou}. Given this setup, we can write the routing problem between these two markets as
\begin{align*}
\text{maximize} \quad & y + z_2 \\ 
\text{subject to} \quad & \{(x,y) \ | \ \sqrt{(R +x)(R' - y)} \geq \sqrt{R R'}, x,y \geq 0\},\\
\quad & \{(z_1, z_2) \ | \ p_0 z_1 \geq z_2, z_2 \leq V_0, z_1 , z_2 \geq 0\}, \\ 
\quad & x + z_1 \leq D. 
\end{align*}
Notably, the routing problem can be formulated by decoupling the variables across the two markets and simply appending the trading set of the limit order to the usual routing problem in the presence of CFMMs. Since the objective function is linear, and all the constraints are convex, this optimization problem remains a convex problem, and can be readily solved with off-the-shelf libraries such as \texttt{cvxpy}. We plot the optimal solution of this problem as $D$, the total input size, varies in both the cases where there is and is not a limit order in Figure \ref{fig:pigou-limit}. Interestingly, we see that this curve exactly matches the modified forward exchange function depicted in Figure \ref{fig:modified-forward-exchange}. Therefore, we see that the modified forward exchange function of a CFMM in the presence of limit orders can be recovered as the solution to a convex optimization problem.

% \begin{figure}
%     \centering
% \includegraphics[width=0.65\linewidth]{figures/limit-order/pigou-limit.png}
%     \caption{Optimal routing for the Pigou network with limit orders.}
%     \label{fig:pigou-limit}
% \end{figure}

\subsection{General Numerical Example} 
We now provide a general numerical example that shows how the routing problem can be scalably and efficiently solved for markets with arbitrarily many limit orders. For this, we adapt an example from \cite{angeris2022optimal}
 in which a user wants to liquidate an amount $s$ of a asset into another asset, which yields a utility we denote $u(s)$. In this example, there are five CFMMs trading between three assets, along with two limit orders at different prices and volumes. We list the market types, fee parameters, and the volumes in Table \ref{tab:market-attributes}. The `fee parameter' of the limit orders is set to $1$, implying that any input trade against the limit order is executed exactly provided the limit order's constraints are met. 

\begin{table}[t]
    \centering
    \begin{tabular}{c | l | c | c}
        \hline
        \textbf{Market} & \textbf{Market type} & \textbf{Fee parameter} $\gamma_i$ & \textbf{Reserves/Volumes} \\
        \hline
        1 & Geometric mean, $w = (3,2,1)$ & 0.98 & (3, .2, 1) \\
        \hline
        2 & Product & 0.99 & (10, 1) \\
        \hline
        3 & Product & 0.96 & (1, 10) \\
        \hline
        4 & Product & 0.97 & (20, 50) \\
        \hline
        5 & Sum & 0.99 & (10, 10) \\
        \hline 
        6 & Limit order at price $0.5$ & 1 & 40 \\ 
        \hline 
        7 & Limit order at price $0.2$ & 1 & 20 \\
        \hline 
    \end{tabular}
     \caption{Market attributes.}
    \label{tab:market-attributes}
\end{table}

\paragraph{Formulation and utility.} The user wants to tender an amount $s \geq 0$ of asset $1$ into the maximum possible amount of asset $3$. The CFMMs are identical to the example in \cite{angeris2022optimal}, but there are two limit orders at prices between asset $3$ and asset $1$, at price-volume pairs $(0.5, 40)$ and $(0.2, 20)$, respectively. We introduce variables $z_1 = (z^1_1, z^2_1)$ and $z_2 = (z_2^1, z_2^2)$ for these limit orders, and construct the net trade $\Psi$ by summing up the total trades in the CFMMs and the limit orders. The user tenders an amount $h^{\text{init}} = t e_1$, where $e_1 \in \mathbf{R}^3$ is the unit vector. The utility of the user is therefore $\Psi_3$ provided $\Psi + h^{\text{init}} \geq 0$ and is negative infinity otherwise. We denote the maximum amount of asset $3$ received for tendering $s$ units of asset $1$ by $u(s)$. 

\begin{figure}[h]
    \centering
    \includegraphics[width=0.8\textwidth]{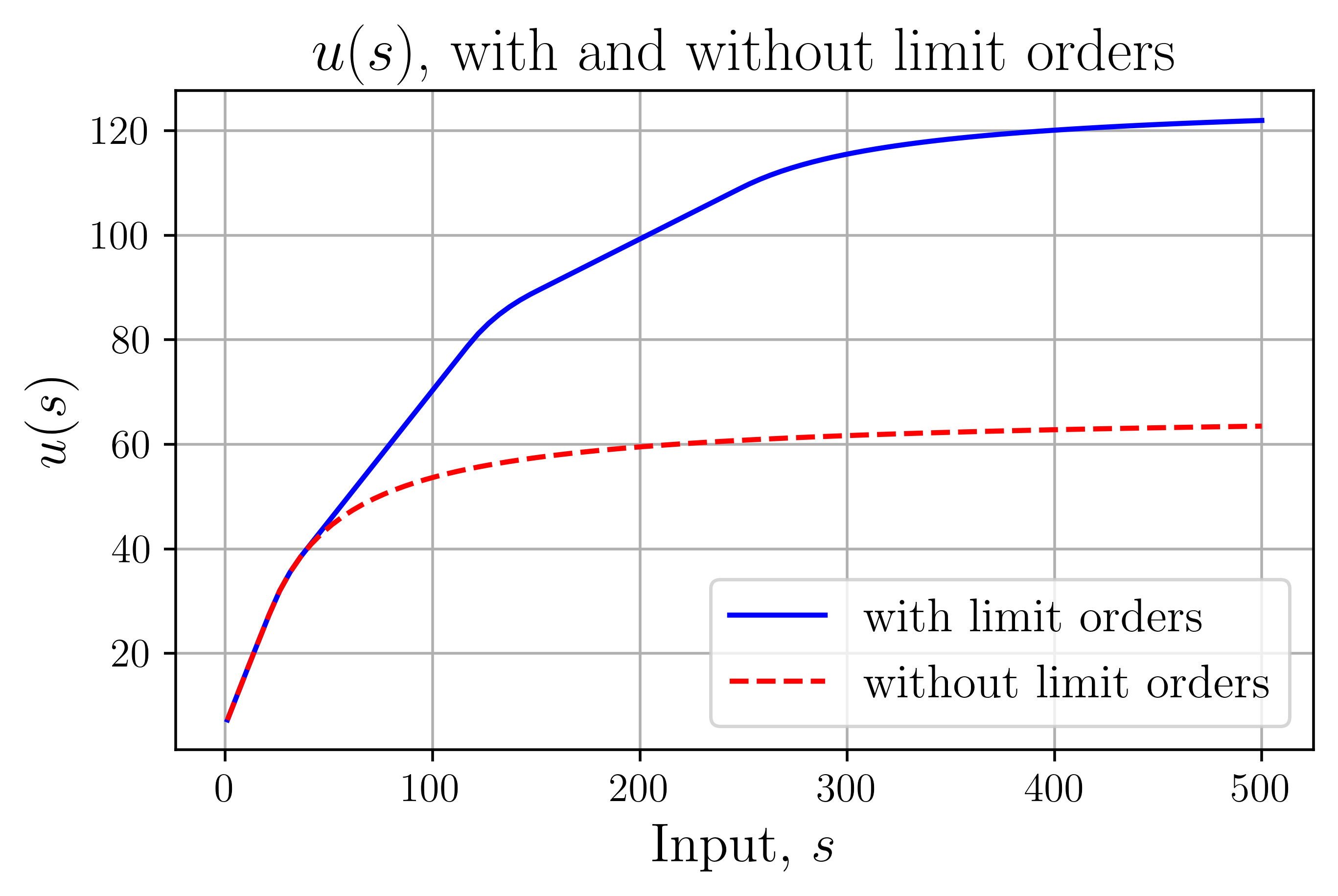}
    \caption{Optimal amount of asset $3$ received, $u(s)$, for the multi-asset limit order network as $s$, the amount of asset $1$ tendered, varies in the presence of limit orders.}
    \label{fig:u-multi-limit}

    \vspace{1cm} % Adjust vertical spacing between figures

    \includegraphics[width=0.8\textwidth]{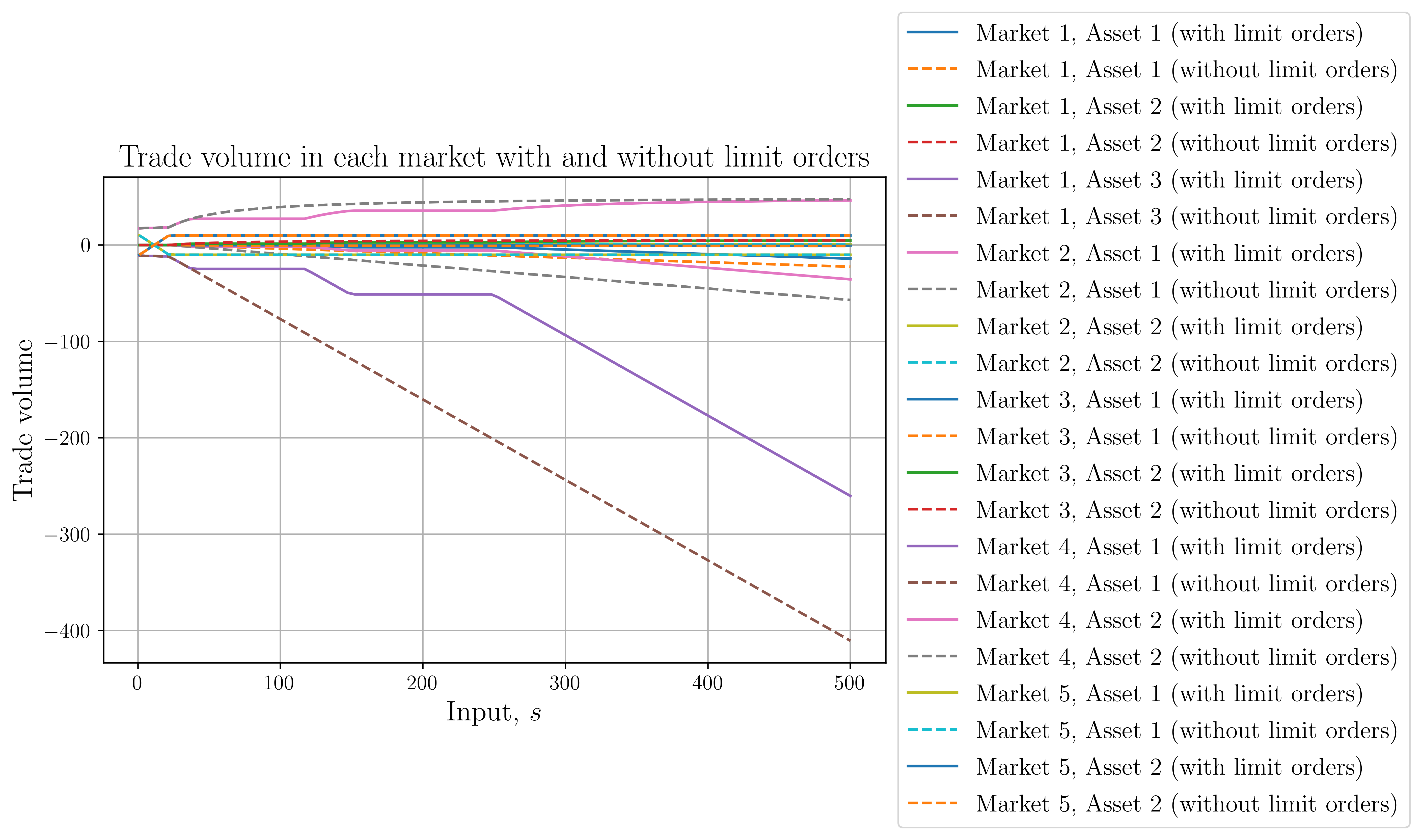}
    \caption{Optimal trades as $s$, the amount of asset $1$ tendered, varies in the presence of limit orders.}
    \label{fig:trades-multi-limit}
\end{figure}

\paragraph{Results.} We vary $s$ over a range from from $0$ to $500$ and solve the optimal liquidation problem using \texttt{cvxpy}. We plot the optimal output $u(s)$, with and without the limit orders, in Figure \ref{fig:u-multi-limit} and observe several interesting phenomena that go beyond optimal routing in conventional CFMMs. First, it can be seen that the optimal output in the presence of limit orders, as the input trade gets large, consumes both the limit orders of net size $60$. Therefore, the additional liquidity provided by the limit orders is used by the problem. Second, we see the presence of `kinks' in the optimal output curve as the limit order prices are reached in the optimization problem and as we move from one order into the other. This demonstrates that the problem is able to gracefully move through the orders and that trading sets of limit orders can be combined with conventional CFMM trading sets. 

In Figure \ref{fig:trades-multi-limit}, we plot the corresponding trades in the CFMMs with and without the limit orders. Here, we observe interesting phenomena regarding the usage of the CFMMs. In particular, we see that certain markets (such as the solid purple line, depicting the trade of asset $1$ in market $4$), are affected in a nonintuitive fashion due to the limit orders. There are points of transition (for example, near $t = 250$) in the optimal trades in the market, as the limit order prices are achieved.

Jointly, our numerical results show that the optimal routing problem with limit orders can be scalably solved using off-the-shelf packages and demonstrate behavior that emerges only when limit orders and CFMM liquidity are combined together. 
\newpage 

\clearpage % Ensures the figures are placed before moving to the next section

\section{Optimal Liquidations: Routing Over Time}\label{sec:temporal-hook}
In this section, we consider the example of a user who wants to liquidate a large inventory of a risky asset $A$ into another asset $B$ (for instance, liquidating a large amount of ETH into USDC).
Traditional CFMMs face a challenge when executing orders of this kind and quote users worse prices due to the potential of a small set of strategic and informed users knowing something the overall market does not.
This problem was studied in~\cite{paradigm-twamm}, in which the authors proposed a new class of CFMMs, called time-weighted average market makers, or TWAMMs, which split up orders into infinitely many infinitely small pieces and execute them over time. TWAMMs have the benefit that the user's order can be `virtually' executed over the time horizon while paying gas costs only once up front. These market makers can be implemented as hooks in Uniswap V4.   

Users and routers have another option, however. If they know auxiliary information about the volatility of the risky asset, they can perform an `optimal' liquidation that improves the user's output over the TWAMM strategy. Concretely, knowing the volatility allows the user to construct a model of the mispricing between the price quoted by the CFMM and the price on an external market. Most CFMMs in practice collect fees from any trader that interacts with them. These fees constrain the set of profitable trades that a sophisticated arbitrageur can make between the external market and the CFMM, since the arbitrageur will only synchronize the prices between these markets until the trade is profitable while accounting for fees. If a router knows the volatility of the risky asset and has a model of the mispricing process, it can use this information to profitably liquidate the inventory in times when the mispricing is in the user's favor, which can improve upon the TWAMM benchmark.

We formulate the optimal liquidation problem as a dynamic program that uses volatility information to liquidate the user's inventory over time. The dynamic program is constructed by first formulating a model of the mispricing process between the price quoted by the CFMM and the price quoted on an external market. The basic tradeoff the user faces when implementing the sophisticated strategy over the TWAMM strategy is that the mispricing process might give additional information about when trading is in her favor, but implementing these trades incurs gas costs. Correspondingly, we formulate an objective function that captures this tradeoff and use numerical methods to solve the program.

\subsection{Model Formulation}
The user holds $D$ units of the risky asset $A$ and wants to trade them into another asset $B$ on a constant product market maker (CPMM) with trading function $\varphi(R, R' , x, y) = (R +x)(R' -y) = K$ and liquidity constant $L = \sqrt{K}$.
Naturally, trading the entire amount in a single swap will cause the user to incur price slippage on the CPMM, which will reduce her output.
Therefore, she considers splitting her trade over a time horizon of length $T$. Assume that each time step $t$ in this time horizon corresponds to a single block of a blockchain, which allows her to make a trade $\Delta_t$. We assume that the user does not run into transaction ordering issues that are common in most modern blockchains, but can always execute her trade after a sophisticated arbitrageur resets the price to a fee bound; we describe the process by which this happens shortly. Trading at any time on the blockchain incurs a fixed gas cost $g$.

The CPMM has reserves $(R_t, R_t')$ and the instantaneous price quoted by the market can be computed as $p_t = R_t'/R_t$. We assume there is an external market trading the same assets, which quotes a price $p_t'$, which generally will be different from the CPMM's quoted price. For exposition, we define the forward exchange function of the CPMM at a \emph{particular} price $p_t$ as $G(\Delta, p_t)$ (this is analogous to defining the forward exchange function for a particular set of reserves). 

When liquidating her position, therefore, the user faces a tradeoff. She can decide to uniformly split her trade up over the time horizon and incur the gas cost $g$ only once, by using a virtual order in a TWAMM. Alternatively, if she has a model for how the price of the external market deviates from the CPMM's price, she can decide to trade at times preferential to her in exchange for incurring gas costs every time she trades. Motivated by this, we first formalize how the mispricing between the two prices evolves. 

\paragraph{The mispricing process.} We define the log mispricing process analogously to~\cite{milionis2023automated} as $z_t = \log \left(p_t' / p_t \right)$.
The CPMM has lower and upper fee bounds, $(\gamma_-, \gamma_+)$, which represent the boundary of the no-arbitrage fee interval (\eg~the region where arbitrageurs do not trade against the pool because it is not profitable when accounting for fees).
If the mispricing is greater than the upper fee bound $(z_t > \gamma_+)$, a sophisticated arbitrageur with inventory onchain and on the external market will buy in the pool until $z_t = \gamma_+$ and sell profitably on the external market. If the mispricing is below the lower fee bound $(z_t < -\gamma_-)$, the arbitrageur will sell in the pool until $z_t = -\gamma_-$ and will buy on the external market. We model this by defining an intermediate variable $z_t^*$.
From our non-strategic user's perspective, the resultant \emph{actionable} mispricing process to decide on is after the sophisticated arbitrageurs trade:  
\begin{align*}
    z_t^* = \begin{cases}
        \gamma_+  & \text{if } \ z_t > \gamma_+ \\
        z_t  & \text{if } \ z_t \in [-\gamma, \gamma] \\
        -\gamma_-  & \text{if } \ z_t < -\gamma_-
    \end{cases}
\end{align*}
Now, suppose the user makes a trade $\Delta_t$ in the pool at time $t$, canonically defined as selling the risky asset. Then, the mispricing process evolves as a perturbed geometric brownian motion (\cite{oksendal2013stochastic}) as
\begin{equation*}
    z_{t+1} = z_t^* \exp\bigl((\mu - \sigma^2/2)  dt + \sigma \sqrt{dt} \cdot \epsilon_t + \tilde{J}_t(\Delta_t)\bigr),
\end{equation*}
where $\mu$ and $\sigma$ are the mean and standard deviation of the mispricing process, $dt$ is the time interval between discrete steps (the blocktime of the blockhain), $\epsilon_t \sim \mathcal{N}(0,1)$, and $\tilde{J}_t(\Delta_t)$ is the jump process from the user's trade, defined as
\begin{align}\label{eq:jump}
    \tilde{J}_t(\Delta_t) = - 2\log \left(1+ \Delta_t \cdot \frac{\sqrt{p_t}}{L}\right).
\end{align}
The quantity $\tilde{J}_t(\Delta_t)$ represents excess price impact from trading against the CPMM. In particular, when $\Delta_t = 0$, i.e. there is no trade, the term $J_t(\Delta_t)$ vanishes. For any positive $\Delta_t$, the jump process is negative, because the user is selling into the pool. We derive the particular functional form of equation \eqref{eq:jump} in Appendix~\ref{app:jump}.

\paragraph{The TWAMM strategy.}
An alternative to paying gas costs repeatedly in an ad-hoc liquidation is to place a \emph{single} virtual order that fractionally divests the user's entire position over a predetermined time horizon, $T$.  In practice, a TWAMM achieves this by continuously executing infinitesimal trades against the CPMM, and whenever an external agent (e.g., an arbitrageur) interacts with the contract, the pool state is ``lazily'' updated to account for all the virtual trades that would have occurred up to that point.  

To illustrate the expected value the user receives from a TWAMM over all stochastic price paths, consider first a minimal \emph{discrete} variant of the TWAMM with equally spaced time-steps within $[0,T]$.  The user splits her inventory $D$ into $T$ pieces of size $\tfrac{D}{T}$, and at each step $t=1,\ldots,T$ liquidates $\tfrac{D}{T}$ units of the risky asset.  Let $p_t'$ be the external price at step $t$; the arbitrage-adjusted price in the pool is thus $p_t' \exp (z_t^*)$. Because the user pays a \emph{single} gas fee $g$ to initiate the TWAMM order, the total expected output in this discrete approximation is
\[
   \mathbf{E}\Biggl[\sum_{t=1}^T p_t' \exp (z_t^*)  \tfrac{D}{T}\Biggr] \;-\; g,
\]
where the expectation is taken over all possible realizations of the future price path (and corresponding mispricing corrections by arbitrageurs).  By design, the TWAMM strategy guarantees that the user continuously trades near the external market price within the fee bounds, thereby mitigating slippage over the entire horizon in addition to paying gas costs exactly once.

\paragraph{Optimal liquidation: state transitions.}
A more sophisticated strategy the user may consider is using the external signal of the mispricing process to trade.
Instead of uniformly splitting her inventory over time, the user considers strategically trading when the mispricing is in her favor.
We formulate this as a Markov decision process (MDP) whose state at time $t$ is given by $S_t = (I_t, z_t)$, where $I_t$ is user's remaining inventory ($I_t = D - \sum_{t' =1}^{t} \Delta_{t'}$) and $z_t$ is the mispricing at time $t$. The action is the trade made by the user, $\Delta_t$. The state transitions of the MDP are given by 
\begin{align}
    I_{t+1} &= I_t - \Delta_t, \notag \\
    z_{t+1} &= z_t^* \exp\bigl((\mu - \sigma^2/2)  dt + \sigma \sqrt{dt} \cdot \epsilon_t + \tilde{J}_t(\Delta_t)\bigr).
    \label{eq:state-transitions}
\end{align}
These transitions encode the fact that the user's trade reduces her inventory and causes a jump in the mispricing, which is reset in the next block by a sophisticated arbitrageur. Constructing this model requires having approximate information about the mean $\mu$ and volatility $\sigma$ of the mispricing process. This information could be provided by hooks that read historical data into the pool.
\paragraph{Optimal liquidation: reward.} To formulate the user's reward in the MDP, we posit that the user is rewarded for trading when the mispricing is in her favor. Ultimately, she wants to clear her inventory, and therefore, she cannot simply wait forever to liquidate. Further, because she trades in an ad-hoc fashion, she incurs gas costs every time she places an order. We construct a reward model that balances her incentive to trade with a running inventory cost and gas costs. We define the reward function accordingly:
\begin{align*}
    R(S_t, \Delta_t) =  G(\Delta_t; p_t' \exp(-z_t)) - G(\Delta_t; p_t') - g \cdot \mathbbm{1}_{\{\Delta_t > 0\}} - \xi I_t 
\end{align*}
where $g$ is once again the gas cost for executing a swap, and $\xi$ is a parameter that encodes the inventory cost. The function $G(\Delta; p)$ encodes the fact that we are considering the forward exchange function at a particular instantaneous price $p$. The first two terms of this reward function encode the fact that the user is rewarded for trading when the mispricing is in her favor. These terms compute the excess output received from trading at the post-arbitrage price relative to trading when the CPMM is pegged at the external market's price. The second term encodes the fact that the user pays gas cost every time there is a trade. Note that this is distinct from the TWAMM strategy, where gas is only paid once for the virtual order. The third term encodes the running inventory cost. The user's objective is to maximize the cumulative expected reward over the time horizon $T$, subject to the state transition dynamics in equation \eqref{eq:state-transitions}.

\begin{figure}[t]
\centering
\begin{minipage}[b]{0.47\textwidth}
    \centering
    \includegraphics[width=\textwidth]{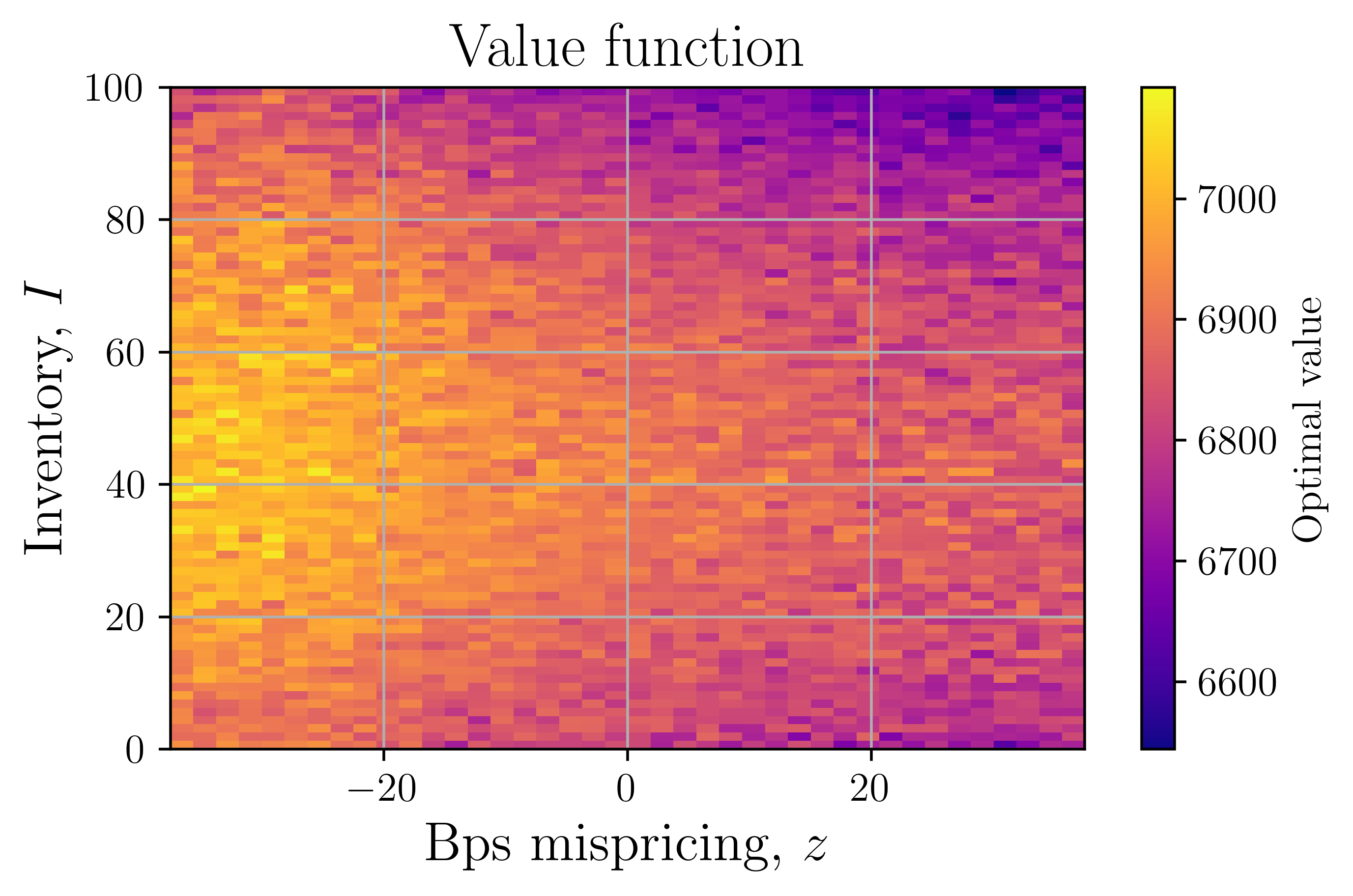}
    \captionof{figure}{Value function.}
    \label{fig:value-function}
\end{minipage}
\hfill
%----- Second figure in minipage -----
\begin{minipage}[b]{0.47\textwidth}
    \centering
    \includegraphics[width=\textwidth]{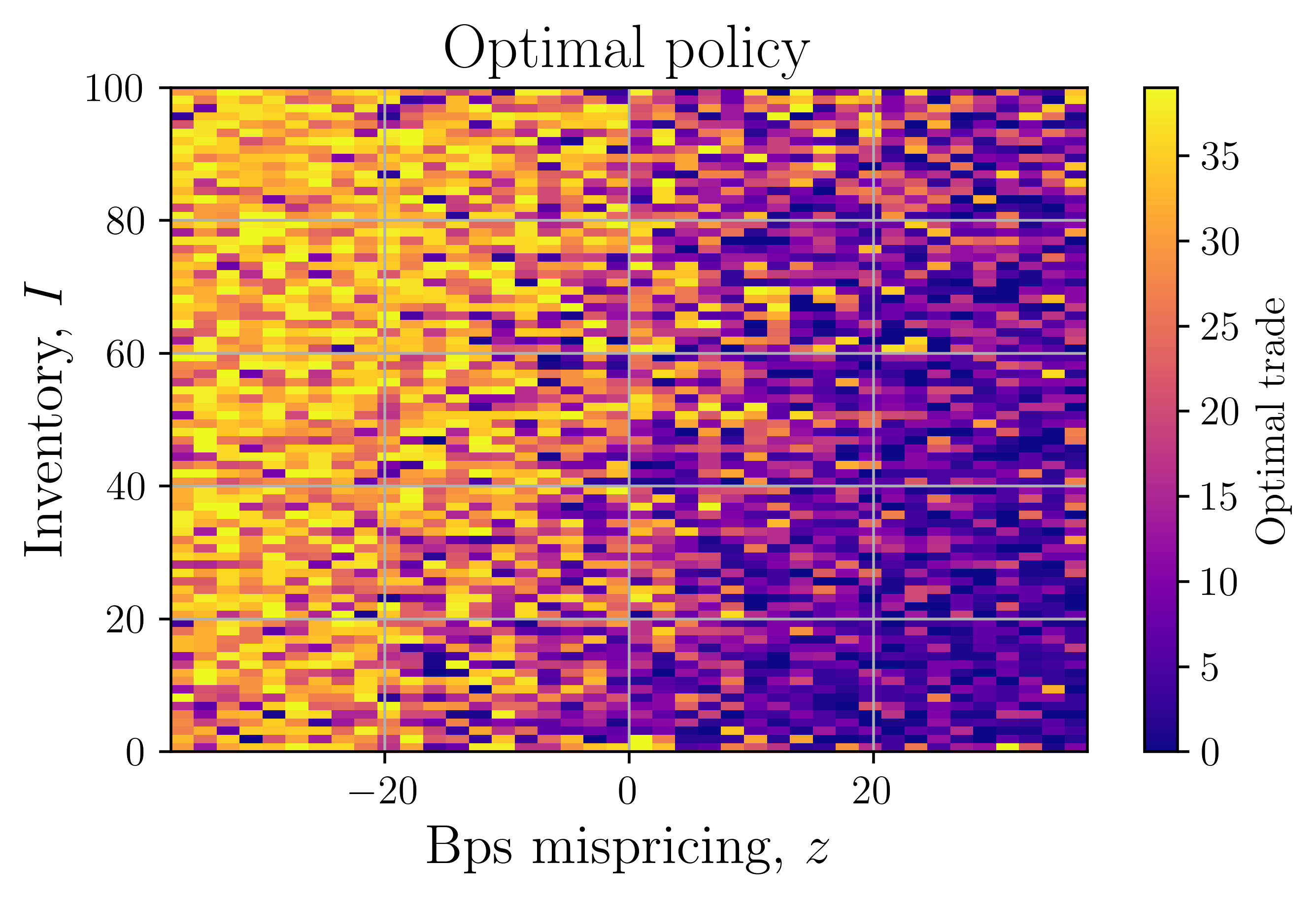}
    \captionof{figure}{Optimal policy.}
    \label{fig:optimal-policy}
\end{minipage}
\caption{Value function and optimal policy for the liquidation problem.}
\end{figure}

\subsection{Results}
We now demonstrate our numerical results by solving the above dynamic program and comparing it to the TWAMM strategy of uniformly splitting the liquidation over time. Solving the dynamic program gives us the value function, which provides the optimal value that can be achieved by starting from any state, and the optimal policy, which provides the optimal trade the user should make at any state. 

\paragraph{Value function and optimal policy.} We numerically solve the dynamic program via a value iteration method for a CPMM with initial reserves $(R, R') = (10^4,5000 \cdot 10^4)$ and fee bounds of $\gamma_+ = \gamma_- =30$bps, over a time horizon of $T = 1000$ blocks, with interblock times of $dt = 1$ second. We set $\mu = 0$ and $\sigma=8$, $g = 2$, $\eta = 0.1$, and $D = 1000$. This simulates a user liquidating a large inventory of ETH into USDC on a standard Uniswap pool with $30$bps fees. We set the discount factor of the dynamic program to be $0.01$. 

Figure \ref{fig:value-function} shows the computed value function. As expected, the optimal value is largest when the mispricing is negative (in the user's favor). Further, when the mispricing is not favorable, the value function is negative, indicating that the uses is penalized for trading in this regime. Figure \ref{fig:optimal-policy} shows the optimal policy, which shows that the user should liquidate larger amounts when the mispricing is negative. Jointly, these results show that the intuitive strategy of trading when the mispricing is in the user's favor and holding back when it's not can be formulated as a dynamic programming problem. 

\begin{figure}[t]
    \centering
    \begin{subfigure}[b]{0.45\textwidth}
        \centering
        \includegraphics[width=\textwidth]{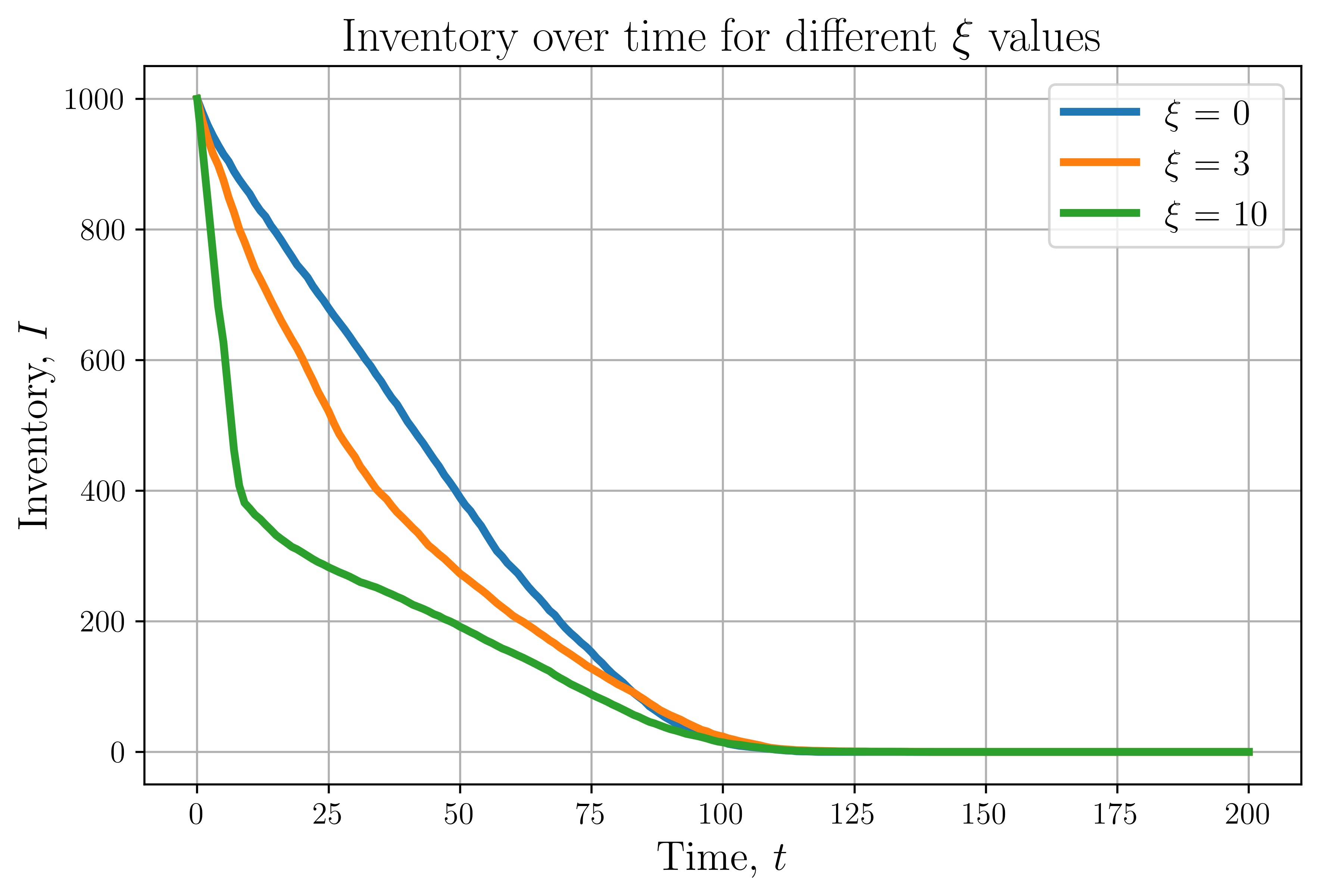}
        \caption{Inventory over time for varying values of $\xi$.}
        \label{fig:xi-plot}
    \end{subfigure}
    \hfill
    \begin{subfigure}[b]{0.48\textwidth}
        \centering
        \includegraphics[width=\textwidth]{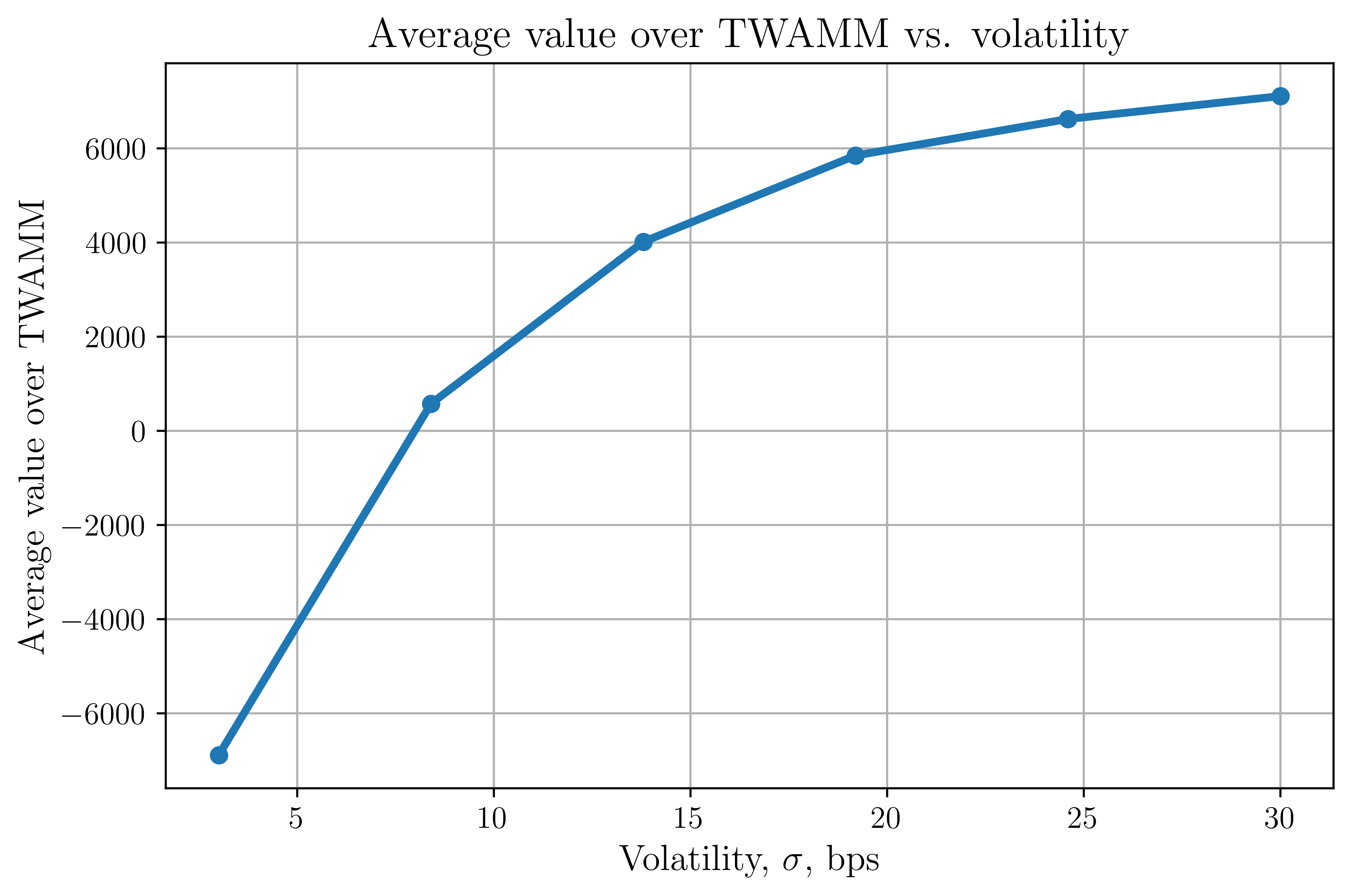}
        \caption{Comparing optimal liquidation to TWAMM.}
        \label{fig:comparison-twamm}
    \end{subfigure}
    \caption{Inventory over time and dollars liquidated over TWAMM for selling $1000$ units of a risky asset priced at $\$5000$ over $1000$ blocks}
    \label{fig:combined}
\end{figure}

\paragraph{Inventory over time.} In Figure \ref{fig:xi-plot}, we plot the inventory over time following the optimal policy for various values of the tradeoff parameter $\xi$, for $T= 200$, $D = 1000$, $(R, R') = (10^5, 5000 \cdot 10^5)$, $\gamma_+ = \gamma_- = 30$ bps, and $g = 2$. We see that this parameter allows us to control how aggressively the user liquidates their inventory. When $\xi = 0$, the user incurs no cost for holding onto the inventory, and therefore, the liquidation occurs slowly. When $\xi = 10$, the user incurs a large inventory cost, and therefore, nearly $60\%$ of the inventory is liquidated within the first $10$ blocks. This matches the intuition and provides a lever for the optimal liquidation problem to make use of the user's preferences. 

\paragraph{Improvement over TWAMM.} In Figure \ref{fig:comparison-twamm}, we compare the output of the optimal liquidation to the TWAMM strategy described above. We plot the average value achieved over the TWAMM for a liquidation with $T = 1000$, $D = 1000$, $(R, R') = (10^5, 5000 \cdot 10^5)$, $\gamma_+ = \gamma_{-} = 30$ \text{bps} and $g = 2$. This corresponds to liquidating a large inventory of a risky asset like ETH into an asset like USDC. The average value we calculate is the net USDC amount received from the optimal liquidation over the TWAMM, accounting for gas costs. Our results demonstrate that the liquidation can achieve levels of output greater than the TWAMM in regimes of high volatility. This corresponds to selling the risky asset when the mispricing is preferential. 

% We now consider comparing the optimal liquidation strategy to the TWAMM strategy. The user's net output from the optimal liquidation strategy is: 
% \begin{align*}
%     \sum_{t=1}^{T}  G_t(a_t) - \sum_{t=1}^{T} g \cdot \mathbbm{1}_{\{a_t > 0\}}.
% \end{align*}
% Note that in this strategy, the user pays gas every time she interacts with the CFMM. It is natural to ask when the optimal liquidation strategy outperforms the na\"ive strategy.
% In Figure \ref{fig:comparison-twamm}, we plot the percentage improvement in the net output from the optimal liquidation strategy over the TWAMM strategy, versus the volatility of the mispricing process. We see that as the volatility increases, so does the outperformance. This is because the external market signal allows the user to trade in times when the mispricing is in her favor, and not trade when the mispricing has moved against her. This outweighs the excess gas costs. This suggests that hooks that maintain external state and route user trades over time can outperform static strategies. 

% \begin{figure}[t]
%     \centering
%     \includegraphics[scale=0.33]{figures/liquidation/value_vs_volatility.png}
% \caption{Comparing optimal liquidation to TWAMM.}
%     \label{fig:comparison-twamm}
% \end{figure}

% Begin new addition
\section{Routing Through Non-Composable Hooks}\label{sec:non-composable-hooks}

The final class of hooks that we formulate the routing problem for are non-composable hooks, such as Angstrom~\cite{Sorella}.
These hooks have sovereignty over their application-specific state and hold their own liquidity provided by external market makers, and thus cannot be routed through with other CFMMs in a synchronously-composable fashion. Routers must chose to put up their own inventory and subject themselves to non-deterministic execution risk when using these pools to get a pricing advantage on their aggregate route.

The routing problem for these hooks can be modeled as an extension of the CFMM Pigou network introduced Section \ref{sec:limit-orders}. In this scenario, a router must decide how to split a user's trade between a traditional composable CFMM pool and a non-composable hook pool. The fundamental tradeoff in this optimization problem is between execution certainty (or fill risk) and potential price improvement. We formulate two optimization problems related to routing in the presence of noncomposable hooks. First, we set up a mean-variance optimization problem that trades off the risk incurred in exchange for additional output. We demonstrate that this problem is a convex optimization problem for a parametric family of noncomposable hook output functions and show the optimal trades for various forms of the variance.  Second, we calculate the efficient frontier, which allows a user to calculate the minimum risk required in order to reach a threshold return. This also is a convex optimization problem; we demonstrate numerical results that calculate the efficient frontier. 

\subsection{Model Formulation}
Consider a user who wants to trade $D$ units of an asset $A$ into another asset $B$. The user has access to a composable pool, which is a standard constant product market maker, and a non-composable hook, for which we construct a mean-variance model that accounts for additional liquidity provided by the external market makers. Let $\Delta$ represent the units of the trade routed through the non-composable hook, with $D-\Delta$ going through the composable CFMM.

\paragraph{Composable CPMM.} The first market the user has access to is a standard constant product market maker with reserves $(R, R')$, which offers deterministic execution via its forward exchange function 
\begin{align*}
    G_1(\Delta) = \frac{-R R'}{R+\Delta} +R'.
\end{align*}
Routing through this market carries no risk (except for transaction-ordering risk found in all blockchains, which we do not model here). When a user sends a trade through this market, it is executed exactly according to the forward exchange function at the given reserves. 

\begin{figure}[t]
    \centering
    \includegraphics[width=0.5\linewidth]{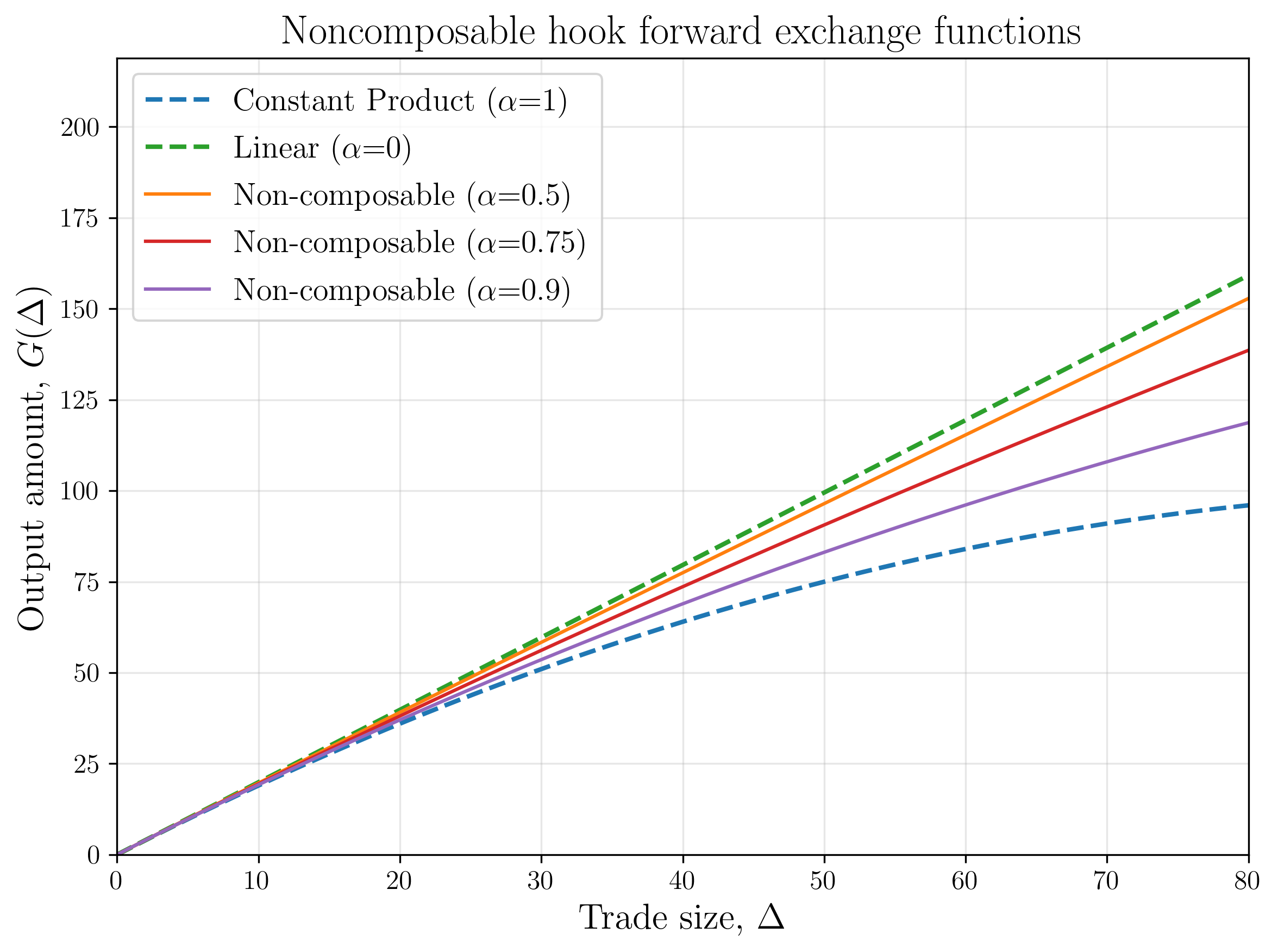}
    \caption{Forward exchange functions of the noncomposable hook for varying values of $\alpha$.}
    \label{fig:parametric-family}
\end{figure}

\paragraph{Non-composable hook.} The second pool presents both an opportunity and a challenge. This pool is provided by hook that allows the user to trade in a fashion that is non-composable with the earlier constant product market maker. In exchange, it potentially provides better prices than the composable CFMM can offer due to the presence of external market makers. Due to the non-composability of this market, the user incurs the risk that their order is not filled while routing. 

The non-composable hook holds reserves $(R_n, R_n')$. To model trading with this hook, we define a parametric family of forward exchange functions (depicted in Figure \ref{fig:parametric-family}) as 
\begin{align*}
    G_2(\Delta) = 2\Delta - \frac{R_n'}{R_n}\Delta^{1+\alpha},
\end{align*}
where $\alpha \in [0,1]$ represents the curvature of the non-composable hook's forward exchange function. When $\alpha = 0$, this function is linear, implying that the hook quotes a constant execution price regardless of swap size. When $\alpha = 1$, the hook behaves like a second order approximation to a constant product CFMM. Intermediate values of $\alpha$ represent the potential for price improvement due to external market makers. This family of functions therefore parametrizes the price improvement that a user might expect to get from the non-composable hook. We use this family of functions as the basis for our mean-variance optimization problem.

\paragraph{Variance.} The non-composable hook's execution is subject to fill risk. While routing through this hook, it is possible that the user's trade does not go through or that the user gets more or less output than that quoted by the output function $G_2(\Delta)$. 
To model this, we introduce a variance term that is a function of the size of the trade through the non-composable hook, $\sigma^2(\Delta)$. When the trade size is large, external market makers may face larger uncertainty while filling the user's trade as opposed to when the trade is small. To formalize this, we assume that the variance is an increasing function of the trade and can take on various functional forms, such as linear ($\beta \Delta$), quadratic, ($\beta \Delta^2)$, or superlinear $(\beta \Delta^{\gamma})$, where $\gamma \in (1, 2)$, and $\beta$ is a scaling parameter. 

\subsection{Results}
We now introduce two optimization problems that yield insight into the interaction of the non-composable hook with the composable CFMM. The first problem is a mean-variance optimization problem that a router can solve, which trades off the additional output from the non-composable hook with the execution risk, modeled by the variance. The second problem computes the efficient frontier, which computes the minimum execution risk that the user must take on to achieve a target combined output from both the CFMM and the hook. 

\begin{figure}[t]
    \centering
    \begin{subfigure}[b]{0.45\textwidth}
        \centering
        \includegraphics[width=\textwidth]{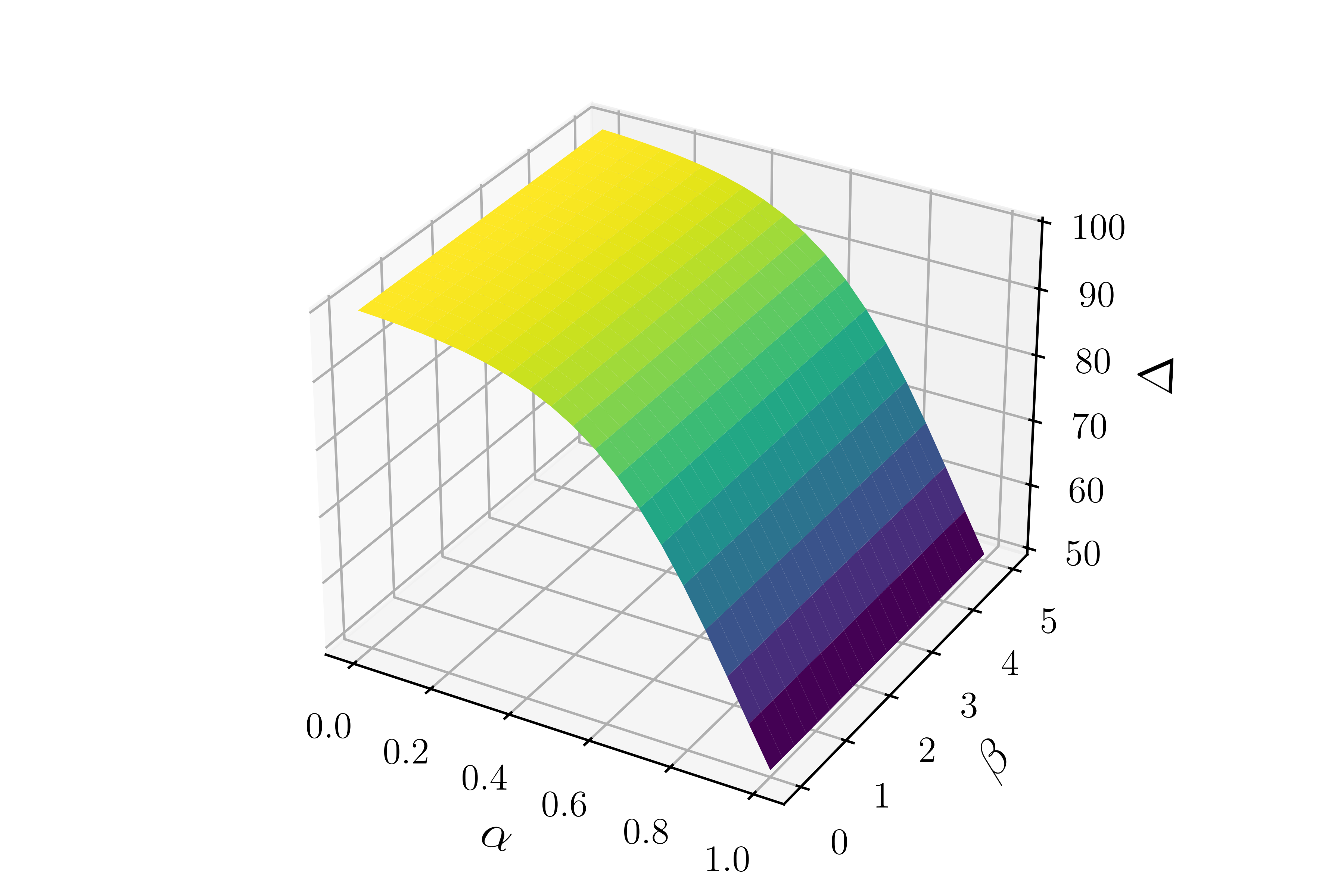}
        \caption{Constant variance}
        \label{fig:constant-variance}
    \end{subfigure}
    %\hfill
    \begin{subfigure}[b]{0.45\textwidth}
        \centering
        \includegraphics[width=\textwidth]{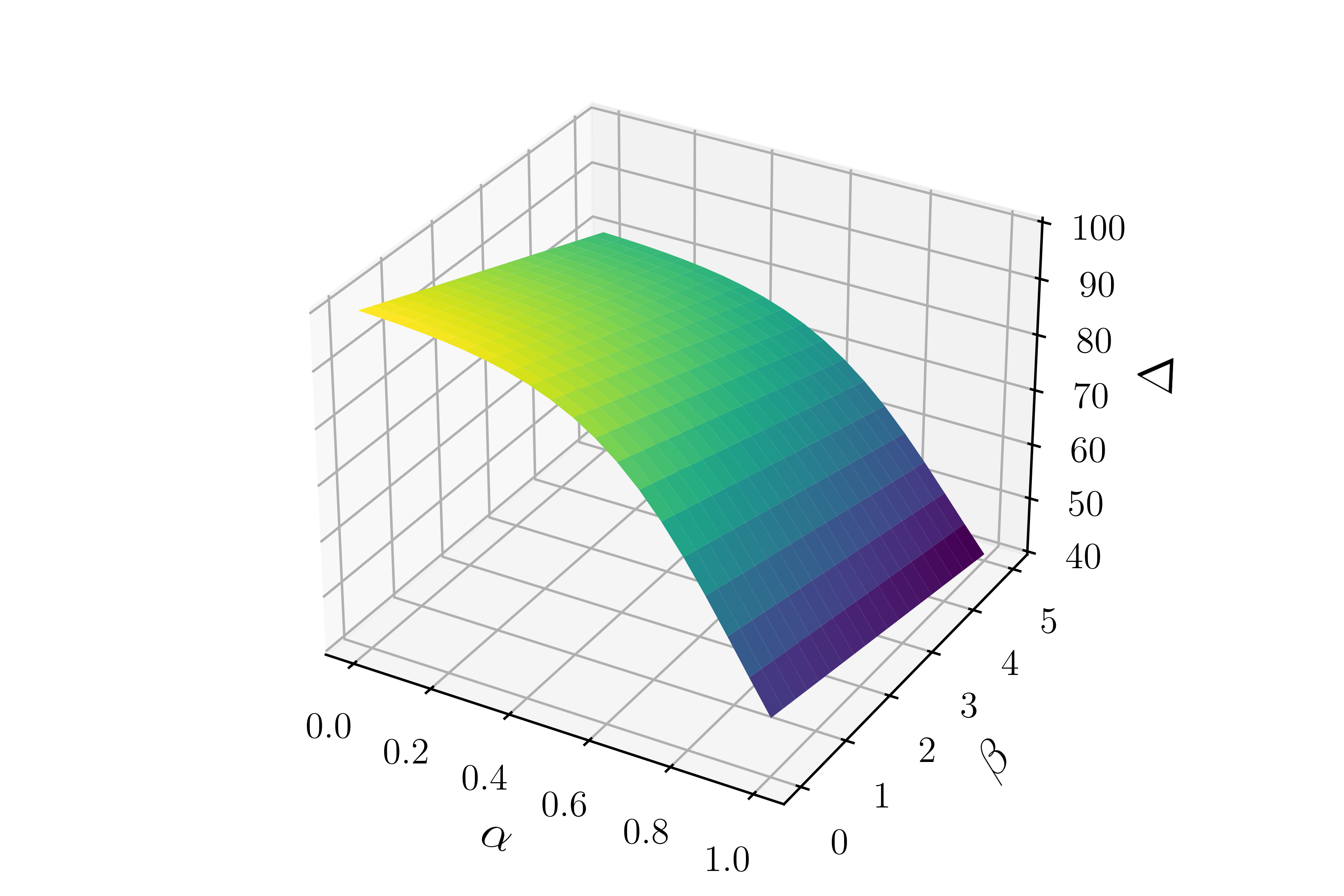}
        \caption{Linear variance}
        \label{fig:linear-variance}
    \end{subfigure}
    
    \vspace{0.5cm}

    \begin{subfigure}[b]{0.45\textwidth}
        \centering
        \includegraphics[width=\textwidth]{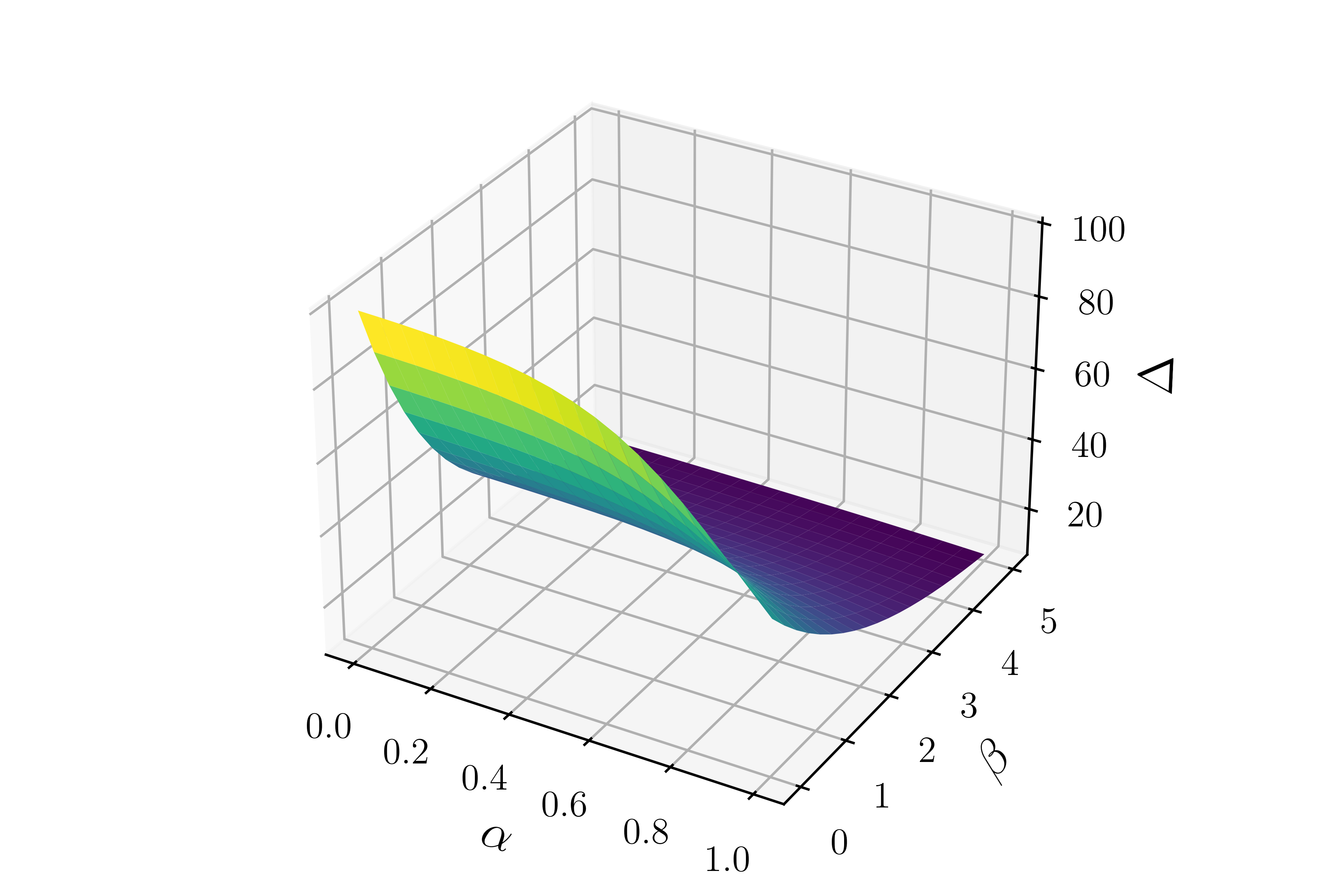}
        \caption{Superlinear variance}
        \label{fig:superlinear-variance}
    \end{subfigure}
    %\hfill
    \begin{subfigure}[b]{0.45\textwidth}
        \centering
        \includegraphics[width=\textwidth]{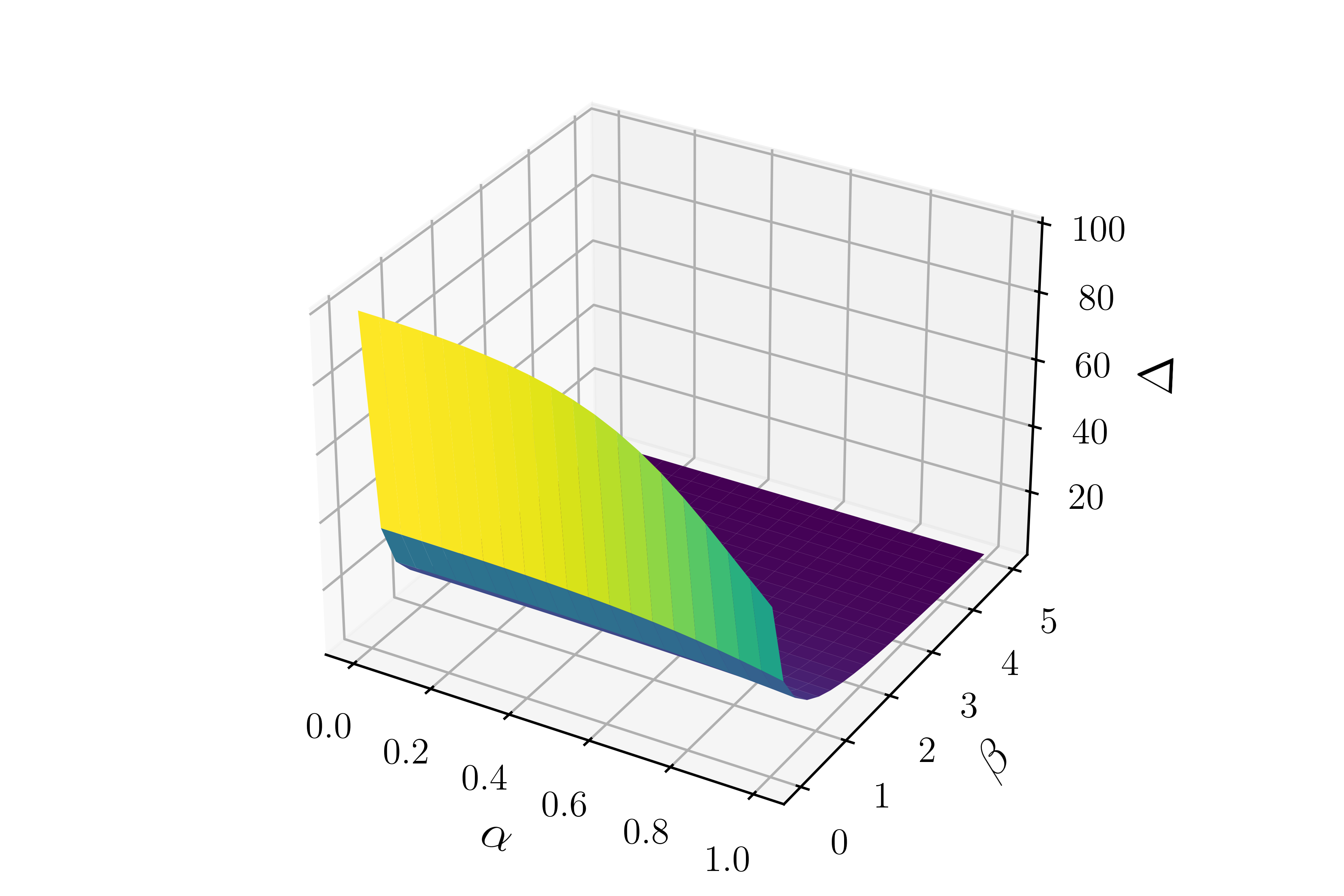}
        \caption{Quadratic variance}
        \label{fig:quadratic-variance}
    \end{subfigure}
    
    \caption{The mean-variance tradeoff for various functional forms of fill risk, $\sigma^2(\Delta)$.}
    \label{fig:mean-variance}
\end{figure}

\paragraph{Mean-variance optimization.} We can formulate the mean-variance optimization problem that encodes the router's risk tradeoff when facing the non-composable hook as
\begin{align*}
    \text{maximize} \quad &  G_1(D-\Delta) + G_2(\Delta) - \lambda\sigma^2(\Delta) \\
    \text{subject to} \quad & 0 \leq \Delta \leq D
\end{align*}
where $\lambda > 0$ is a risk tradeoff parameter. This is a convex optimization problem whenever $G_1$ and $G_2$ are concave functions, as is the case for most realistic CFMMs, and $\sigma^2(\Delta)$ is convex. We solve this problem in \texttt{cvxpy} for constant, linear, superlinear, and quadratic variance terms and plot the optimal trade through the noncomposable hook, $\Delta^*$, as $\alpha$, the parameter that governs the output of the hook, and $\beta$, the parameter that scales the variance, in Figure \ref{fig:mean-variance}. 

We observe several interesting features of the optimal trade. In the case of constant variance (Figure \ref{fig:constant-variance}), we see that the optimal trade is only a function of the non-composable hooks forward exchange function, and in particular of the parameter $\alpha$. The trade through the non-composable hook is maximized for $\alpha = 0$. In the case of linear, superlinear, and quadratic variance (Figures \ref{fig:linear-variance}, \ref{fig:superlinear-variance}, \ref{fig:quadratic-variance}), we observe that the trade through the non-composable hook decreases as the variance gets large. Further, for quadratic variance (Figure \ref{fig:quadratic-variance}), there is a sharp dropoff in the trade through the hook.

\paragraph{The efficient frontier.} We can also use this framework to understand the user's efficient frontier when trading with the non-composable hook (the amount of risk the user must take to achieve a given target level of return). To compute the efficient frontier, we can set a threshold return $\tau$ and ask that the variance, or the risk taken by the user, be minimized, through the following optimization problem:
\begin{align*}
    \text{minimize} \quad & \sigma^2(\Delta) \\ 
    \text{subject to} \quad & G_1( \Delta) + G_2(D - \Delta) \geq \tau \\
   \quad & 0 \leq \Delta \leq D
\end{align*}
Once again, this is a convex optimization problem, as the variance $\sigma^2(\Delta)$ is convex, and the constraint $G_1(\Delta) + G_2(100 - \Delta) - \tau \geq 0$ is convex whenever $G_1$ and $G_2$ are concave functions. We solve this problem in \texttt{cvxpy} for the same choices of $G_1(\Delta)$ and $G_2(\Delta)$ as above. 

\begin{figure}
    \centering
    \includegraphics[width=0.5\linewidth]{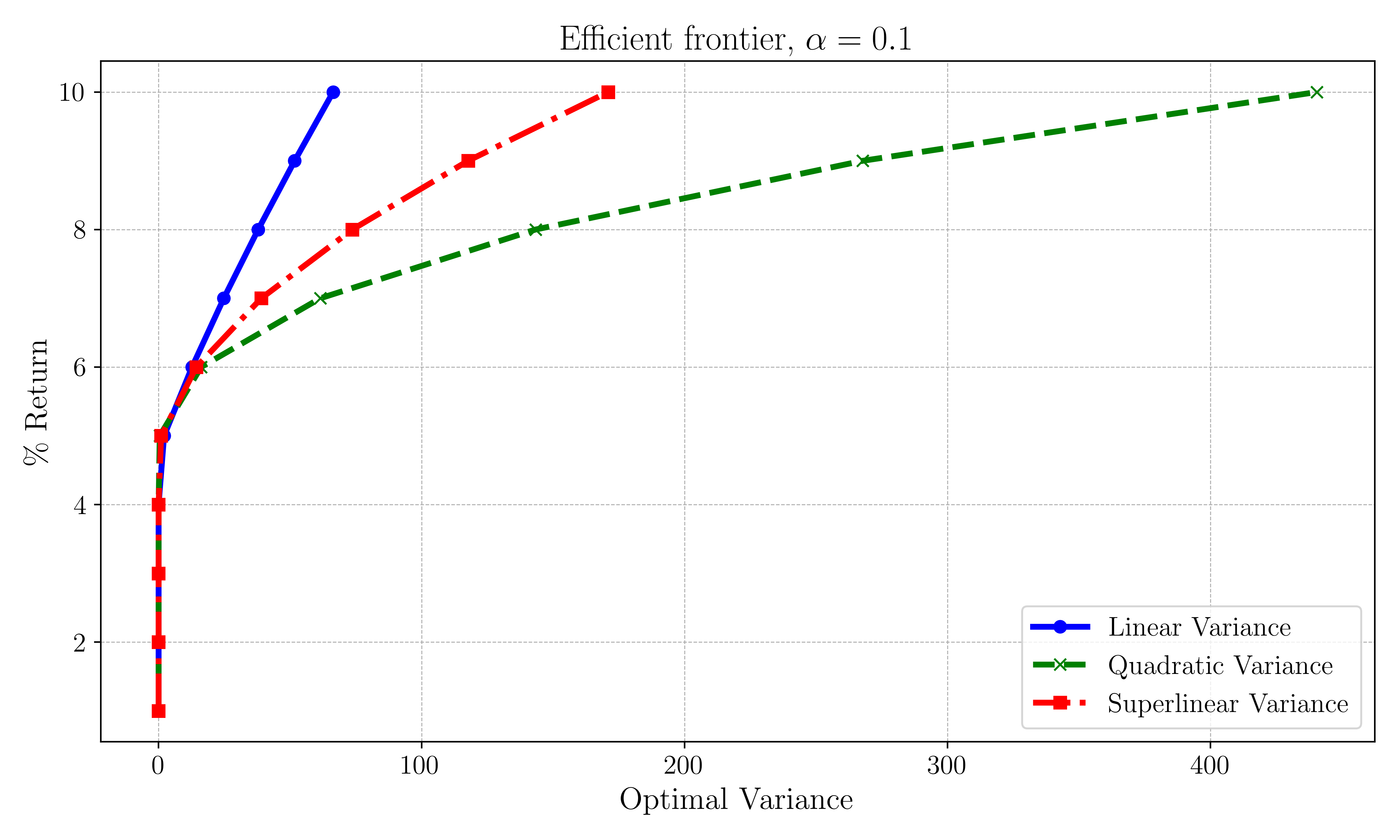}
    \caption{Efficient frontier for the non-composable hook.}
    \label{fig:efficient-frontier}
\end{figure}

We set $\tau$ to range from $0$ to $10$ and plot the user's percentage return versus the optimal variance incurred by the user in Figure \ref{fig:efficient-frontier}, for $\alpha = 0.1, $ in the case where the variance is linear, quadratic, and superlinear for $\beta = 1$. We see that the efficient frontier is pushed further out for variance terms that grow quickly (quadratic, as opposed to linear, for example): the user has to take more risk to achieve the same level of return. This allows us to quantify the tradeoff between increased return and risk that all users face when interacting with noncomposable hooks. 

\section{Conclusion and Future Work}
In this paper, we generalize the optimal routing problem for CFMMs to allow for hooks and consider various settings. First, hooks allow for the implementation of onchain limit orders. We show that limit orders act as concentrated liquidity positions at price ticks of infitesimally small width and demonstrate how to composably route trades between CFMMs and limit orders. Next, we consider the optimal liquidation problem for a user that wants to make a large trade. We formulate this problem as a Markov decision process when the user has access to a mispricing signal with respect to an external market and benchmark the solution relative to a strategy that liquidates uniformly over time, the TWAMM strategy. Finally, we consider venues for trading where users face fill risk in exchange for potential price improvement, called non-composable hooks. We characterize the mean-variance tradeoffs in these markets and demonstrate the efficient frontier for user trades. 

Our results explore the new design space opened by hooks and provide techniques that may be useful for practitioners, motivated by convex optimization and dynamic programming. 
Future work involves creating a general framework for routing in the presence of arbitrary hooks and developing adaptive algorithms that routers can use to improve trade outcomes for users over time. 

\section{Acknowledgments}
The authors would like to thank the Uniswap Foundation for providing a grant for this work and Ciamac Moallemi, Danning Sui, and Theo Diamandis for helpful comments and feedback.
\printbibliography
\appendix
\section{The CPMM Jump Process}\label{app:jump}
A CFMM with trading function $(R+x)(R'-y) = RR' = K$ has a liquidity constant $L = \sqrt{K}$ \cite{adams2021uniswap}.
Suppose that at time $t$, the reserves are $R_t$ and $R'_t$. The corresponding instantaneous price quoted by the CFMM is
\[
p_t = \frac{R_t'}{R_t'} = \frac{L}{R_t^2}
\]
which is equivalent to $R_t = \frac{L}{\sqrt{p_t}}$. After a trade of size $\Delta_t$, the price is updated to
\[
p_{t+1} = \frac{L}{(R_t+\Delta_t)^2}
\]
This implies that the logarithmic jump in price realized from the trade is
\begin{align*}
    \log\left(\frac{p_{t+1}}{p_t}\right) = -2  
    \log \left(\frac{R_t + \Delta_t}{R_t} \right) = -2 \log \left(1 + \frac{\Delta_t}{R_t}\right)  = -2 \log \left( 1 + \frac{\sqrt{p_t}}{L} \Delta_t \right)
\end{align*}
which equivalent to $\tilde{J}_t(\Delta_t)$ in equation \eqref{eq:jump}.

\end{document}